%% file: main.tex
\documentclass[sigconf,authorversion]{acmart}

\AtBeginDocument{%
  \providecommand\BibTeX{{%
    \normalfont B\kern-0.5em{\scshape i\kern-0.25em b}\kern-0.8em\TeX}}}

\setcopyright{acmlicensed}
\copyrightyear{2024}
\acmYear{2024}
\acmDOI{10.1145/3613904.3642865}

\acmConference[CHI '24]{Proceedings of the CHI Conference on Human Factors in Computing Systems}{May 11--16, 2024}{Honolulu, HI, USA}

\acmBooktitle{Proceedings of the CHI Conference on Human Factors in Computing Systems (CHI '24), May 11--16, 2024, Honolulu, HI, USA}
\acmISBN{979-8-4007-0330-0/24/05}

\usepackage{xspace}
\usepackage{subcaption}
\usepackage{enumitem}
\setlist[itemize]{leftmargin=*}

\newcommand{\ie}{i.e.{\xspace}}
\newcommand{\eg}{e.g.{\xspace}}

\newcommand{\qt}[1]{\textit{``#1''}}

\newcommand{\iqr}[1]{$IQR=#1$}
\newcommand{\md}[1]{$MD=#1$}

\newcommand{\rev}[1]{#1}

\newcommand{\tool}{OutlineSpark\xspace}
\newcommand{\tooloutline}{\textit{Outline Panel}\xspace}
\newcommand{\toolcode}{\textit{Notebook Overview}\xspace}
\newcommand{\toolslide}{\textit{Slides Panel}\xspace}
\newcommand{\ckeyword}{\textit{Keyword Extraction}\xspace}
\newcommand{\ctopic}{\textit{Topic Recommendation}\xspace}
\newcommand{\cretrieval}{\textit{Cell Retrieval}\xspace}
\newcommand{\cgeneration}{\textit{Slide Generation}\xspace}

\newcommand{\actor}{Andie\xspace}

\begin{document}

\title{\tool: Igniting AI-powered Presentation Slides Creation from Computational Notebooks through Outlines}

\author{Fengjie Wang}
\affiliation{
  \institution{Sichuan University}
  \city{Chengdu}
  \country{China}
}
\affiliation{
  \institution{The Hong Kong University of Science and Technology}
  \city{Hong Kong SAR}
  \country{China}
}
\email{wangfengjie@stu.scu.edu.cn}

\author{Yanna Lin}
\affiliation{
  \institution{The Hong Kong University of Science and Technology}
  \city{Hong Kong SAR}
  \country{China}
}
\email{ylindg@connect.ust.hk}

\author{Leni Yang}
\authornote{Correspondence authors.}
\affiliation{
  \institution{The Hong Kong University of Science and Technology}
  \city{Hong Kong SAR}
  \country{China}
}
\email{lyangbb@connect.ust.hk}

\author{Haotian Li}
\affiliation{
  \institution{The Hong Kong University of Science and Technology}
  \city{Hong Kong SAR}
  \country{China}
}
\email{haotian.li@connect.ust.hk}

\author{Mingyang Gu}
\affiliation{%
  \institution{Sichuan University}
  \city{Chengdu}
  \country{China}
}
\email{gumingyang@stu.scu.edu.cn}

\author{Min Zhu}
\authornotemark[1]
\affiliation{%
  \institution{Sichuan University}
  \city{Chengdu}
  \country{China}
}
\email{zhumin@scu.edu.cn}

\author{Huamin Qu}
\affiliation{
  \institution{The Hong Kong University of Science and Technology}
  \city{Hong Kong SAR}
  \country{China}
}
\email{huamin@cse.ust.hk}

\renewcommand{\shortauthors}{Wang et al.}

\begin{abstract}
Computational notebooks are widely utilized for exploration and analysis. 
However, creating slides to communicate analysis results from these notebooks 
is quite tedious and time-consuming.
Researchers have proposed automatic systems for generating slides from notebooks, 
which, however, 
often do not consider the process of users conceiving and organizing their messages from massive code cells.
Those systems ask users to go directly into the slide creation process,
which causes potentially ill-structured slides and burdens in further refinement.
Inspired by the common and widely recommended slide creation practice: drafting outlines first and then adding concrete content,
we introduce \tool, an AI-powered slide creation tool that generates slides from a slide outline written by the user.
The tool automatically retrieves relevant notebook cells based on the outlines and converts them into slide content. 
We evaluated \tool with 12 users.
Both the quantitative and qualitative feedback from the participants verify its effectiveness and usability.

\end{abstract}

\ccsdesc[500]{Human-centered computing~Interactive systems and tools}
\ccsdesc[500]{Human-centered computing~Natural language interfaces}
\ccsdesc[300]{Applied computing~Document preparation}

\keywords{computational notebooks, slides generation, outlines, data science}

\maketitle

\input{texfiles/1-v3-intro}
\input{texfiles/2-relatedwork}

\input{texfiles/3-design}

\input{texfiles/4-tool}

\input{texfiles/5-usagescenario}

\input{texfiles/6-userstudy}

\input{texfiles/7-v2-discussion}
\input{texfiles/8-conclusion}

\begin{acks}
We would like to thank the reviewers and all participants in our studies for their valuable input. 
The research was partially supported by the National Natural Science Foundation of China (Grant No. 62172289) and the Hong Kong Research Grants Council (GRF16210722).
\end{acks}



\end{document}

%% file: texfiles/1-v3-intro.tex
\section{Introduction}

Computational notebooks like JupyterLab~\cite{jupyterlab} and Jupyter Notebook~\cite{perez2015project} are widely popular for data exploration and analysis~\cite{kluyver2016jupyter, rule2018exploration}. 
With a block-based interface and dynamic code execution, computational notebooks enable an iterative analysis process in which users flexibly create cells to explore new analysis approaches and directions, inspect intermediate results, and make documentation~\cite{kery2018story,kross2019practitioners,perkel2018jupyter}.
However, such an analysis process often results in lengthy and poorly formatted notebooks~\cite{head2019managing, rule2018exploration}.
This brings challenges when users need to present critical analysis details and results to teammates in collaboration or report to stakeholders such as clients and decision-makers.
Users have to utilize external presentation tools (\eg, Google Slides or Microsoft PowerPoint) to create slides for effective communication~\cite{chattopadhyay2020s, piorkowski2021ai}.
They have to put lots of effort into 1) conceiving the structure and content of their presentations from messy notebooks \cite{piorkowski2021ai, nb2slides}, 2) locating and retrieving relevant cells from notebooks to extract details \cite{wang2023slide4n, rule2018exploration, kross2021orienting}, and 3) making slides that align with the intended presentation structure \cite{nb2slides, brehmer2021jam}.

Researchers have proposed automatic systems to alleviate this tedious and burdensome task.
NB2Slides \cite{nb2slides} automatically generates an entire slide deck based on a prescribed outline following the stages of data science and machine learning lifecycle.
However, this approach is only applicable to those computational notebooks for building machine learning models.
Furthermore, it requires the notebook to have a complete data science workflow and high-quality documentation, conditions that are seldom met in practice~\cite{wang2022documentation, rule2018exploration}.
Slide4N \cite{wang2023slide4n} addresses these issues by permitting users to select cells of interest to generate and refine slides.
However, it assumes that the users have a clear image of the organization and content of their slides and it asks users to begin directly with the process of selecting cells.
In this workflow, users may find it too demanding to hold a clear mental image of the structure of their slides to make the selection.
It can also lead to ill-structured slides and bring additional burdens to refining the organization of the generated slides.
To sum up, these tools do not consider the process of slides ideation---conceiving the structure and content of the slides.
There is a need for additional support to facilitate the slide ideation process and to connect the outcomes of this ideation to the actual generation of slides.

To fill the gap, we present \tool, an interactive and intelligent tool that supports generating slides from computational notebooks based on the outlines written by users. 
Specifically, when using the tool, a user can look through the notebooks and write outlines to ideate the structure of the presentation.
An outline can be a general title indicating the purpose or results of some cells such as ``Data Cleaning", ``Removing Outliers'' and ``Findings about Year Built vs Price''.
The outlines will be later used as the input of an automatic algorithm that can retrieve cells relevant to each item in the outline and generate the corresponding slide.

We decided upon the outline-driven workflow mainly for two reasons: 
1) It is a common practice to craft outlines before creating slides, as evidenced by the prevalence of presentations starting with an agenda or outline slide \cite{bergman2010outline, peng2023slide}.
2) It is also highly recommended since it aids in organizing what to present \cite{li2023ai, reynolds2011presentation, zanders2018presentation, anholt2010dazzle}.
It is worth noting that \tool allows users to manually select cells, and it will generate a corresponding item in the outline and the slide.
Our goal is not to replace the previous workflow of selecting cells for slides creation, instead, we aim to complement the lack of consideration in the slides ideation process. 

Centering around the outline-based workflow, \tool (\autoref{fig:ui}) has a set of functions and interface designs to facilitate slides ideation and generation.
First, for the slides ideation, \tool provides \toolcode (\autoref{fig:ui} (A)) that provides an overview of notebook cells with keywords for reminding users of the content of the notebook.
It further recommends the next item in the outline by leveraging the current outline, the location where the recommendation was requested, and the analysis conducted within the notebook.
Second, for the slides generation, the tool automatically retrieves relevant cells from the notebook based on the outlines and converts them into slide contents in \toolslide (\autoref{fig:ui} (C)).
After the slides are generated, the users can refine their content and organization by modifying the outlines directly.
The notebook and slides are fully linked through outlines, allowing for ideation, generation, and refinement of slides.

To evaluate the effectiveness and usability of \tool, we conducted a user study with 12 participants.
Moreover, we attempt to understand users' preferences between generating slides through writing outlines, which we refer to as the \textit{outline-based} approach, and through selecting cells, which we refer to as the \textit{selection-based} approach.
The feedback obtained from the questionnaires and interview demonstrates that \tool could help participants effectively create desired slides with less effort and is easy to use.
We also observed that most participants showed a preference for outline-based slide generation, because it resonated with their regular practices, and allowed them to concentrate on crafting the narrative of their presentation rather than becoming overly fixated on individual slides.
Overall, participants highly praised the process of creating slides simply by outlines.
Finally, we concluded our research by discussing the lessons learned and potential future directions.

In summary, our contributions in this paper include: 
\begin{itemize}[nosep, left=1.8em]
    \item An outline-based workflow that streamlines the slides ideation and creation process from computational notebooks;
    \item A computational notebook plugin, \tool{}, that assists data scientists in creating presentation slides via outlines;
    \item A user study conducted to evaluate \tool{} and understand user preferences between outline-based and selection-based slides generation from computational notebooks.
\end{itemize}

%% file: texfiles/2-relatedwork.tex
\section{Related Work}

\subsection{Tools for Enhancing Computational Notebooks in Data Analysis}
Computational notebooks have become the most popular programming environment for data scientist~\cite{rule2018exploration,kery2018story}.
With computational notebooks, data scientists can iterate through chunks of code, experiment with different methods, inspect intermediate results, and add documentation during exploration~\cite{kross2019practitioners, kery2018story, perkel2018jupyter, head2019managing}.

While widely adopted, data scientists have encountered many problems in using computational notebooks, which has attracted a large number of researchers in the HCI field to develop tools for enhancing notebooks.
These tools have covered almost every stage of data scientists’ workflow, from code search~\cite{li2021nbsearch,li2023edassistant}, to code management~\cite{head2019managing,kery2017variolite,kery2019towards}, data exploration~\cite{lee2021lux, epperson2022leveraging, raghunandan2022lodestar}, machine learning (ML) model development~\cite{ono2020pipelineprofiler,bhat2023aspirations,munechika2022visual}, and others~\cite{kery2020mage, weinman2021fork, wang2022stickyland}. 
In the domain of code search, for example, NBSearch~\cite{li2021nbsearch} supports semantic search for relevant code cells from a large corpus of notebooks and designs novel visualizations to support interactive exploration of search results.
For code management, Variolite \cite{kery2017variolite} and Verdant \cite{kery2019towards} introduce features like rapid versioning and intuitive visual exploration of code history to assist in the effective management of evolving codes.
In terms of exploration, examples include Lux \cite{lee2021lux} and Solas \cite{epperson2022leveraging}  that are designed to automatically recommend static visualizations for exploring data.
As for ML model development, Pipeline Profiler \cite{ono2020pipelineprofiler} utilizes visual analytics to support the exploration and comparison of machine learning pipelines, so as to improve users’ understanding of the algorithms.

While these works primarily focus on the technical tasks performed by data scientists, a limited body of research addresses the less technical but equally important communication tasks.
Previous studies have emphasized that clear and effective communication is essential for data science workers to align expectations, build trust, and share insights~\cite{li2023ai, muller2019data, zhang2020data, mao2019data}. 
However, there are numerous challenges, such as knowledge gaps, language barriers, and issues with trust-building, which complicate communication~\cite{piorkowski2021ai, kang2021toonnote}.
These findings underscore the importance of improving communication within the data science workflow.
Our study contributes to this emerging area of research by focusing on facilitating the creation of slide decks to effectively communicate analysis results from notebooks. 
We will delve into this topic further in \autoref{sec:rw2}.

\subsection{Communication Support for Computational Notebooks} \label{sec:rw2}

With the growing complexity of data analysis work, multiple collaborators with different backgrounds are usually involved in the same analysis \cite{zhang2020data, kim2016emerging, kang2021toonnote}, and they need to frequently communicate with each other to move forward \cite{donoho201750, kross2021orienting, wang2021much, chevalier2018analysis}. 
However, computational notebooks, which represent the technical work of data scientists, are often lengthy, disorganized, and interspersed with interim notes \cite{guo2012software, tabard2008individual, head2019managing, rule2018exploration}. 
It’s difficult for other collaborators to understand, which in turn hinders communication \cite{head2019managing, piorkowski2021ai, kang2021toonnote}.

To address this critical problem, some researchers aim to curate the notebook before sharing.
Adam et al. \cite{rule2018aiding} proposed a technique that controls the visibility of cells using hierarchy to aid the high-level comprehension of notebooks.
Code Gathering Tools \cite{head2019managing} help find, clean, recover, and compare versions of code and generate more readable, cleaner notebooks. 
However, code is still the main content of notebooks, which is hard to understand \cite{rule2018aiding, wang2023slide4n}.

Another thread of research improves the readability of notebooks by facilitating the creation of documentation that provides explanations and findings in notebooks.
For example,
Themisto \cite{wang2022documentation} applies deep-learning algorithms to generate several kinds of documentation (\eg, process and reference) for code in computational notebooks. However, it only provides a start of the sentence as a prompt and asks users to manually document findings.
To alleviate the burden, InkSight \cite{lin2023inksight} allows users to sketch on the charts they created for analysis to indicate data subsets of their interests and automatically generates the documentation of findings from data.
However, it is not always suitable to use the notebook with documentation as the communication medium.
In scenarios such as reporting to stakeholders, the content of notebooks should be filtered and reorganized to make slide decks for presentation, which demands much manual effort.

Recently, some works have attempted to close the gap between data analysis and presentation by supporting slide generation directly from computational notebooks.
RISE~\cite{RISE} allows users to designate the role of each cell within a notebook, categorizing them as slides, sub-titles, elements to skip, and so on. 
Slides are then automatically generated based on these predefined roles. 
Although straightforward, this method requires a manual configuration process.
In contrast, Notable~\cite{li2023notable} seeks to automate this process to some extent by offering an on-the-fly plugin. 
It auto-generates the chart findings and embellishments to help data analysts create slides during the analysis phase. 
Unlike Notable which supports slide creation when analyzing data, some research tries to support the slide generation after data analysis.
For example, NB2Slides~\cite{nb2slides} employs a prescribed outline to distill notebook contents into templated slides. 
While it generates a slide deck with one click, the rigid outline constrains data scientists from freely telling their stories and makes it challenging to adapt to diverse scenarios. 
Furthermore, its demand for high-quality documentation is often impractical in real-world settings.
Slide4N~\cite{wang2023slide4n} mitigates these limitations by allowing users to select individual notebook cells for slide generation. 
However, this method demands users maintain a clear mental image of their presentation structure to make the selection, which can result in disorganized slides and additional effort in refining the organization.
In summary, both NB2Slides and Slide4N overlook the slide ideation process, which is crucial for shaping the presentation structure and content.
In line with the research of supporting slide generation after data analysis, \tool considers how to facilitate the ideation process and connect it to the creation of the slides. 
It facilitates slide creation by enabling users to craft customized outlines while automatically generating the corresponding slides.

%% file: texfiles/3-design.tex
\section{Design Goals}

Our tool is designed to support the slide ideation process in creating presentation slides for communicating essential analysis details and results from computational notebooks. 
Informed by relevant literature, we derive the following design goals.

\textbf{G1: Support the slides creation process guided by outlines.}
Designing a presentation using outlines is a widely recognized good practice \cite{li2023ai, reynolds2011presentation, zanders2018presentation, anholt2010dazzle}. 
They serve as a guide for slide creation and are commonly used, as evidenced by the prevalence of presentations starting with an agenda or outline slide~\cite{bergman2010outline, peng2023slide}.
Thus, the tool should facilitate users in structuring outlines and the creation of the slides centered around the outlines.

\textbf{G2: Retrieve relevant cells based on the outlines.}
Computational notebooks often suffer from disorderly cell execution and loose connections between cells~\cite{head2019managing, subramanian2020tractus, wenskovitch2019albireo}. 
However, creating a slide typically relies on a set of tightly interconnected cells. 
Following the outlines, data scientists need to identify these cells within the notebook, which is tedious and time-consuming. 
Thus, the tool should automatically retrieve relevant cells from the notebook by identifying the connections between cells and the outlines.

\textbf{G3: Automate slides creation from relevant cells.}
Presentation slides typically feature human-readable, self-explanatory, and concise content, such as titles, bullet points, and charts/tables \cite{wang2023slide4n, sefid2021slidegen, zanders2018presentation}.
Cells relevant to the outlines serve as the foundational materials for creating such slides, typically, the users need to manually extract and summarize the key information from them~\cite{nb2slides, rule2018exploration, kross2021orienting}, and arrange them on slides. 
Such a process is often burdensome and time-consuming.
Therefore, the tool should facilitate this process with automation to simplify the creation of slides.

\textbf{G4: Support easy refinement of generated slides.}
Allowing users to exert some control over the slide generation process can serve as a means of refining the generated slides to better meet individual preferences and requirements \cite{nb2slides, heer2019agency, HAIworks}, and mitigate the limitations of automated methods \cite{blohm2018comparing,zukerman2001natural, li2023ai}.
Therefore, the tool should support the refinement of the generated slides with a suite of easy-to-use interactions, such as updating slides via outlines, adjusting cells used in slide generation, and modifying generated slides.

\textbf{G5: Recommend outline candidates based on the notebook.}
Crafting an effective outline that covers the main sections or topics, is a fundamental step in creating organized and coherent slides \cite{li2023ai, reynolds2011presentation, zanders2018presentation}. 
It takes time for users to consume the content of the notebook cells and go back and forth between notebook cells and the outline to draft it clearly.
Although there is no fully automatic algorithm to create an outline that aligns well with the intention of the users, partial automation is possible by recommending the next item in the outline.
To increase the efficiency of drafting outlines, we propose that the tool can recommend items of the outline according to previous items that are written by the users and the analysis in the notebook.

\textbf{G6: Assist with quick recall of the notebook.}
Conceiving the structure of the slides requires users to first identify crucial information from the notebook.
However, computational notebooks often suffer from lengthiness, poor formatting, and a focus on code~\cite{guo2012software, tabard2008individual, head2019managing, rule2018exploration}.
Readers, including oneself in the future, usually lack the interest to thoroughly read such documents to grasp the notebook's essence at a high level~\cite{rule2018aiding}. 
Thus, the tool should provide an overview of the notebook appropriately to enable a quick recall of the entire analysis.

%% file: texfiles/4-tool.tex
\section{\tool} 
In this section, we first present an overview of \tool (\autoref{sec:overview}). 
Then we introduce the interactive modules and computational modules of \tool (\autoref{sec:interactivemodule} and \autoref{sec:computationmodule}, respectively).

\subsection{System Overview} \label{sec:overview}

\tool is developed based on the aforementioned design goals. 
It aims at assisting users in structuring and creating presentation slides from computational notebooks based on the outlines written by the users. 
The system architecture consists of two components: the interactive modules and the computational modules. 
The interactive modules decide the user interface and interaction designs, while the computational modules support the functioning of interactive modules in the back-end.

\begin{figure*}[tb]
  \centering
  \includegraphics[width=\linewidth]{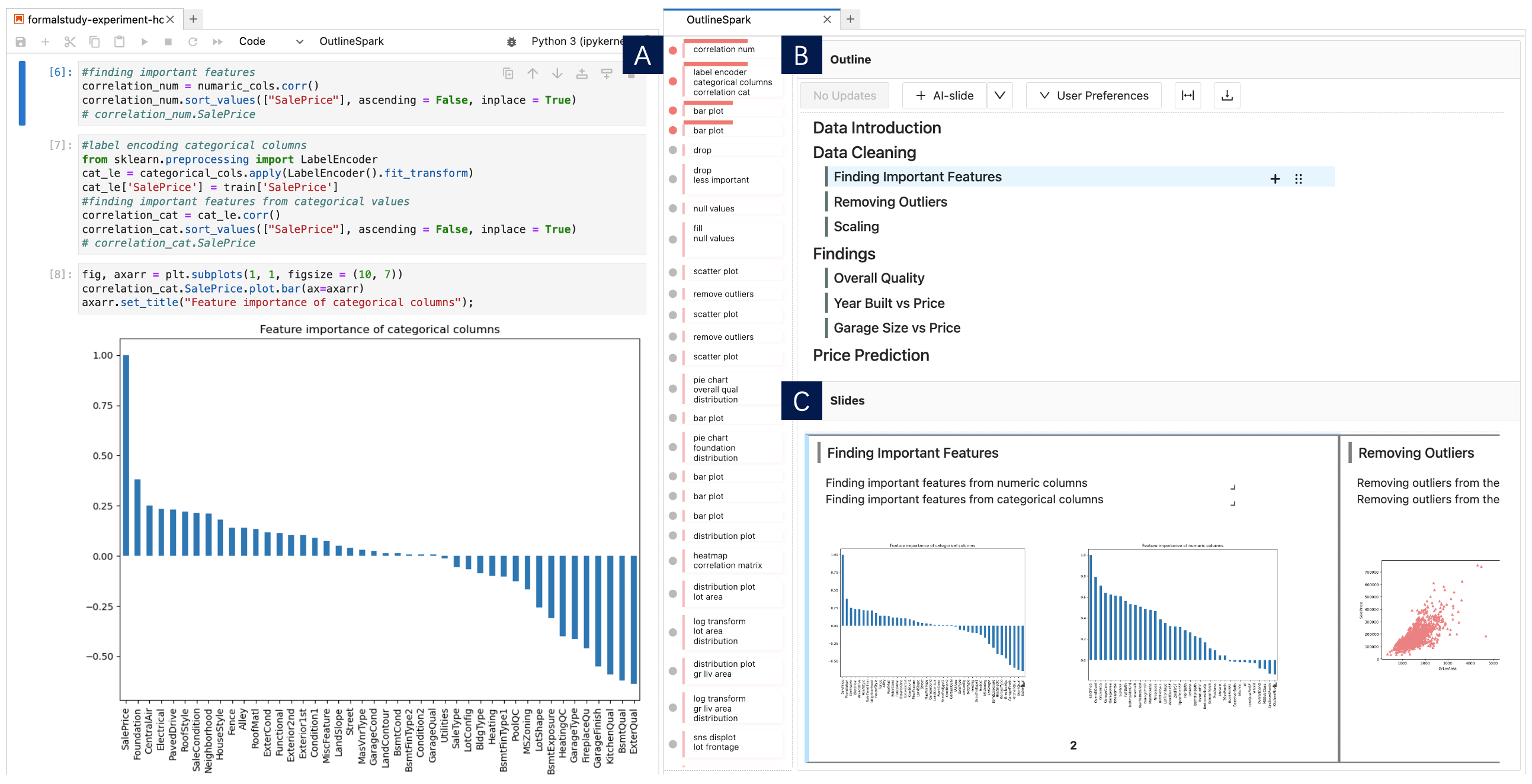} 
  \vspace{-6mm}
  \caption{
\tool is an interactive AI-powered JupyterLab plugin designed to facilitate the creation of presentation slides from computational notebooks using outlines. 
The user interface of \tool comprises three interconnected components:
a \toolcode (A) that provides a visual summary of the notebook cells by relevant keywords,
a \tooloutline (B) that enables users to craft outlines and initiate the generation of slides, and
a \toolslide (C) that renders the generated slides and offers users the flexibility to refine and customize them.
} 
  \Description{The figure presents the user interface of OutlineSpark, integrated into a notebook environment. The interface is divided into two sections: on the left, the notebook itself, and on the right, the OutlineSpark plugin. The plugin consists of three main components:
(A) Notebook Overview: This component displays a vertical column of cards, each representing a cell in the notebook. The height of each card represents the content length of its respective cell. Dots on the left of the cards indicate the selection status (e.g., pink for selected cells). Bars at the top of the cards show the correlations between cell content and the outlines. Tooltips of each card provide summaries of the cell contents.
(B) Outline Panel: This component enables users to create and manage outlines for slide generation. The panel includes a toolbar for slide generation, layout selection, customization options like bullet point detail, and downloading slides as PowerPoint files for further refinement. Below the toolbar, a text editor allows for crafting hierarchical outlines as topics and subtopics, and for adjusting the hierarchy and order of these items.
(C) Slides Panel: This component shows auto-generated slides based on the outline. Users can edit slide content, including text, charts, and tables, and adjust the layout by rearranging elements.}
  \label{fig:ui}
\end{figure*}

As shown in \autoref{fig:ui}, \tool can be positioned side by side with the notebook window, which is implemented as a JupyterLab extension.
There are three interactive modules:
(a) \tooloutline (\autoref{fig:ui} (B)), which allows users to craft outlines (G1, G5) and trigger slides generation (G2, G3); 
(b) \toolcode (\autoref{fig:ui} (A)), which provides a visual summary of the code and Markdown cells in the notebook by keywords (G6). This component also enables users to exclude or include a cell in a slide by selection (G4);
and (c) \toolslide (\autoref{fig:ui} (C)), which renders the generated slides and allows users to refine and customize them (G3, G4).

To support the interactive modules of \tool, we have designed four computational modules: 
(a) \ckeyword, responsible for extracting keywords from notebook cells which are then displayed in \toolcode to assist in summarizing the notebook's content (G6);
(b) \ctopic, which extracts topics from the notebook content, and based on these extracted topics and the current outline, recommends the next outline item for users in \tooloutline (G5);
(c) \cretrieval, which retrieves relevant cells for each item in an outline (G2);
and (d) \cgeneration, designed for extracting essential information from notebook cells to generate slide titles, bullet points, charts, and tables on slides (G3).
In the following sections, we introduce these modules in detail.

\subsection{Interactive Modules} \label{sec:interactivemodule}
This section presents three interactive modules in the front-end user interface to support users in creating presentation slides from computational notebooks via outlines (G1).

\begin{figure*}[tb]
  \centering
  \includegraphics[width=\linewidth]{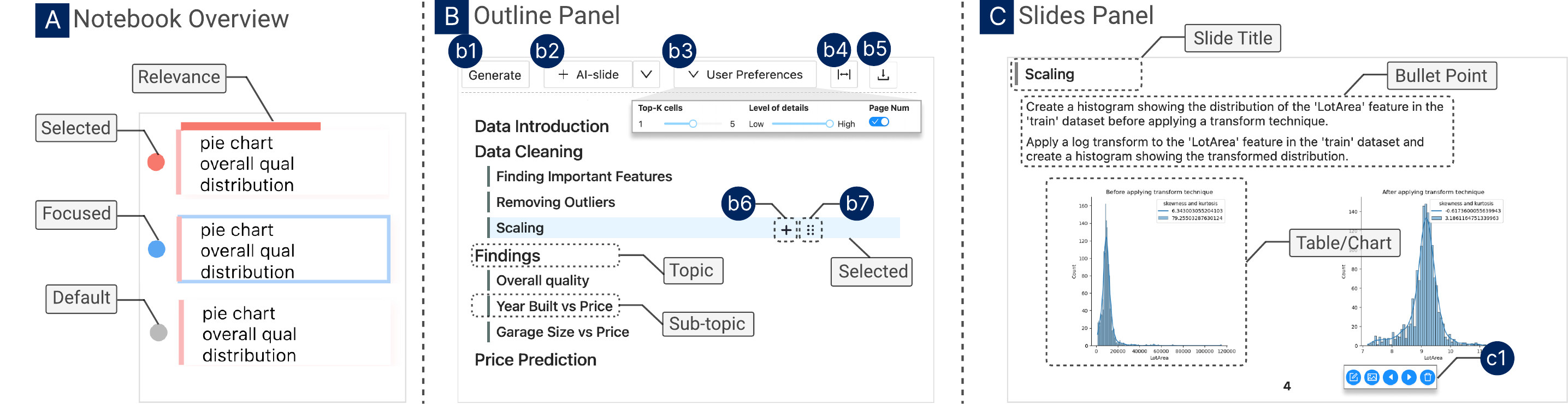} 
  \vspace{-5.5mm}
  \caption{
  This figure illustrates the interactive components of \tool, including (A) \toolcode, (B) \tooloutline, and (C) \toolslide.
  (A) explains the visual encodings of \toolcode.
  (b1)-(b7) allows users to generate slides, adjust slide generation options and parameters, edit outlines, and etc.
  (C) showcases the content of a slide, and (c1) enables users to refine the generated slides.
  } 
  \Description{The figure presents details of three interactive modules of OutlineSpark, labeled as A, B, and C.
(A) Notebook Overview: This component presents a concise overview of the notebook, displaying cells as cards with keywords. The height of each card represents the content length of its respective cell. Cards are arranged in the order of the cells in the notebook, from top to bottom. Dots on the left of the cards indicate the cell's status: default (gray), focus (blue when hovered over), or selected (pink for cells selected for slide generation). Bars at the top of the cards show the correlations between cell content and the outlines.
(B) Outline Panel: This component comprises a toolbar (top) and a text content editor (bottom). The toolbar's five main functions are marked b1-b5. Button b1 ('Generate') initiates slide generation. Button b2 ('+AI-slide') is for manual slide creation. Dropdown box b3 enables adjusting the number of retrieved cells, bullet point detail level, and slide numbering. Button b4 is for slide presentation, and Button b5 allows downloading slides as PowerPoint files. Below the toolbar, a text editor allows for crafting hierarchical outlines as topics and subtopics, distinguished by font size, indentation, and vertical gray bars. Example topics like 'Data Introduction' and 'Data Cleaning' are shown, with 'Data Cleaning' including subtopics such as 'Finding Important Features'. The "Scaling" subtopic was selected, and buttons b6 and b7 on the right indicate the functionality for Outline editing. Users can easily adjust the level and order of outline items by clicking the buttons b7 or by using drag-and-drop interaction.
(C) Slides Panel: This component displays auto-generated slides of the crafted outlines, allowing users to customize them. Slides, arranged from left to right, include titles, content, and visual elements like charts or tables. Users can edit slides, modify charts and tables, and adjust layout by rearranging content elements.}
  \label{fig:components}
\end{figure*}

\textbf{\toolcode}.
As shown in \autoref{fig:components} (A), \toolcode concisely displays the notebook cells by keywords for a quick overview to assist users in drafting outlines (G6).
Each notebook cell is represented by keywords displayed on a card with a pink left border. 
\rev{The height of a card increases with the amount of content in the cell}, and the cards are arranged from top to bottom in accordance with the order of cells in the notebook.
This card-based visual design aims to maintain consistency with the cell-based design in JupyterLab throughout the interface.
Inspired by Slide4N \cite{wang2023slide4n} and NB2Slides \cite{nb2slides}, cells are used to generate slides in \tool.
A dot is positioned on the left side of each card, with its color to indicate three possible states of a cell, namely, \rev{default (gray), focused (blue, indicating the mouse hovering over a card), or selected (pink, indicating a card chosen for slide generation).}
The pink bar on top of the card is designed to help users check the automatically retrieved cells by \cretrieval, which reflects the relevance of the cell to the \rev{selected outline item in \tooloutline (\autoref{fig:components} (B), when an outline item is clicked or edited by the users)}, the more relevant the wider the bar.
The users can double-click on a card to bind or unbind the cell to the selected outline item, which allows them to easily refine the cells used to generate slides (G4).
Several interactions are provided to facilitate seamless navigation between the notebook and \tool.
When the users click on a card, the notebook window jumps to the corresponding cell and vice versa.
Furthermore, hovering over a card triggers a tooltip that provides a quick overview of the cell's content, including code snippets and any associated tables or charts.

\textbf{\tooloutline}.
After a quick overview of the notebook cells through \toolcode, users can move to \tooloutline to draft the outline.
We design \tooloutline (\autoref{fig:ui} (B)) to facilitate users' planning and creating presentation slides using explicit outline (G1).
\tool supports writing a hierarchical outline including topics and sub-topics which are differentiated by font size, indentation, and a vertical gray bar. 
For example, sub-topic outline items are represented with a smaller font size and indented with a vertical bar on the left.
\rev{Users can add an outline item by either pressing ``Enter'' on the keyboard or clicking the button on the right of each item (\autoref{fig:components} (b6)).}
Additionally, they can easily adjust the level and order of outline items by clicking the button or drag and drop interactions on the right of each item (\autoref{fig:components} (b7)).
To get topic recommendations, users can press the space bar. 
When doing so, a list of recommended topics will appear below the selected outline item \rev{for users to choose from.}
After completing the outline, they can click the button at the top of \tooloutline (\autoref{fig:components} (b1)) to prompt \tool to generate a slide deck based on the outline.
If the users are not satisfied with the current outline, they can further edit them.
In addition to generating slides via outlines, \tool also allows users to select cells of interest for slide generation, by double-clicking on the cards in \toolcode (\autoref{fig:components} (A)).

At the top toolbar of \tooloutline, users can change the slide generation options and parameters (G4).
Specifically, 
they can manually create some slides by clicking ``+AI-slide'' (\autoref{fig:components} (b2)), 
adjust the number of cells retrieved (\ie, top-K relevant cells) for generating a slide (\autoref{fig:components} (b3)), 
control the level of detail in the generated bullet points (\autoref{fig:components} (b3)), 
include page numbers on slides (\autoref{fig:components} (b3)), 
stretch slides for presentation purposes (\autoref{fig:components} (b4)), 
\rev{and download the current slides as .pptx  for further customization (\autoref{fig:components} (b5)).}

\textbf{\toolslide}.
After generating the slides, users can proceed to make further refinements on \toolslide (G4).
\toolslide (\autoref{fig:ui} (C)) displays the generated slides in a left-to-right manner and offers a range of accessible interactions to empower users in refining and customizing the slides to meet their preferences.
Inside each slide, users can edit any bullet point \rev{using markdown-based grammar}.
They can insert, resize, or remove plots and tables.
Furthermore, they can adjust the layout by dragging and dropping slide contents (\eg, bullet points and charts).
\tool further provides different layout templates, such as title slides and slides with one or two-column content.
Users can click the ``down arrow'' on the \tooloutline (\autoref{fig:components} (b2)) to select a template.
In terms of the general structure, users can manage slide order, and delete or restore slides (\autoref{fig:components} (c1)). 
Any changes on the slides that affect the outline, \eg, modifications to slide titles, addition, deletion, or reordering of slides, are seamlessly synchronized with the \tooloutline (\autoref{fig:components} (B)).

\tool provides interactive linking between all the interactive modules, allowing users to easily trace back to the cells used to generate the slides.
When the user clicks on an outline item in \tooloutline (\autoref{fig:ui} (B)), 
\toolslide (\autoref{fig:ui} (C)) automatically scrolls to the corresponding slide; 
\toolcode (\autoref{fig:ui} (A)) highlights the retrieved cells (horizontal bar on top of a card) and selected cells (pink dots on the left side of the card), and scrolls to the first selected cell. 
\rev{Notably, when clicking on a slide in \toolslide, \tooloutline and \toolcode respond in a similar manner.}
To help the users track their modifications, \tool highlights adjusted outline items (\eg, modification, addition, deletion, reordering, or binding new cells) in pink.
\rev{
For more advanced slide editing and beautification, users can export the slides in the .pptx format to modify them in Microsoft PowerPoint by clicking the download button (\autoref{fig:components} (b5)).
}

\subsection{Computational Modules} \label{sec:computationmodule}
This section describes the four computation modules that support the interactive modules. 
Considering the remarkable capabilities of LLMs (\eg, GPT-3.5 \cite{openaigptmodel}) in understanding and generating both natural language and code \cite{kasneci2023chatgpt, wei2022emergent, wei2022chain}, we leverage LLM \rev{(\ie, gpt-3.5-turbo-16k with temperature set to 0 for more consistent outputs) to assist users in creating the outline and converting them into slides}.
It is worth noting that 
\rev{(1) \tool's capacity for notebook length is limited by the employed LLM. According to the common length of data science notebooks \cite{rule2018exploration}, 
we opt for 16k tokens to accommodate these notebooks within the context length limits of the LLM;}
(2) All the prompts in computational modules are designed based on existing successful prompt engineering experiences \cite{liu2022design,openaigptprmpt}.
Further details about the prompts can be found in the supplementary material.

\textbf{\ckeyword}.
Users often lack interest in delving into complex code when attempting to understand the notebook at a high-level~\cite{rule2018aiding}.
To reduce the effort, \tool provides all the notebook cells to the LLM and instructs it to summarize the input (\ie, the code) of every notebook cell into at most 5 keywords and ranks them by how representative they are of the cell's content.
The keywords are then rendered explicitly in \toolcode (\autoref{fig:components} (A)), allowing users to quickly grasp the complex code by keywords (G6).

\textbf{\ctopic}. 
To ease the burden of drafting outlines, we instruct the LLM to generate contextually appropriate topic recommendations for users (G5).
Before recommendation, \tool provides the notebook cells to LLM and prompts it to extract topics and relevant sub-topics from these cells as the candidate topics set.
To ensure a user-friendly experience, \tool adopts a request-by-user design for topic recommendations, avoiding unnecessary interruptions. 
When requesting, \tool instructs the LLM to pick the top 10 relevant topics from the candidate topics set based on the current context of the current outline. 
The context depends on where the users make the request. 
As shown in \autoref{fig:components} (B), if requested at the sub-topic level, \tool treats the outline items of the parent topic (including itself) as context; if requested at the topic level, \tool treats all the current topic level outline items as the context.

\textbf{\cretrieval}. 
Finding relevant cells within a cluttered notebook can be burdensome for users, to simplify this process, \tool automatically maps notebook cells to user-created outline items (specifically, those items at the lowest level of the hierarchical outline).
This involves transforming the hierarchical structure of outlines into flat outline units, each comprising two components: the outline item and its corresponding context. 
The context indicates the higher-level topic to which the outline item belongs, providing valuable contextual information for precise mapping.
Subsequently, \tool provides the LLM with all outline units and notebook cells to ensure it makes informed mapping decisions.
\tool then instructs the LLM to map each outline unit to a maximum of 5 notebook cells, taking into account the semantic relevance between the outline unit and the input of notebook cells. 
The retrieved cells are also assigned semantic relevance scores ranging from 0 to 1. 
To inform users of the mappings, the retrieved cells and their relevance scores are then visualized in \toolcode (\autoref{fig:components} (A) Relevance).

\textbf{\cgeneration}. 
Creating slides from notebook cells requires considerable effort, \tool automates the process to ease the burden (G3).
Inspired by Slide4N \cite{wang2023slide4n}, \tool tasks the LLM to convert every cell into one bullet point to summarize the content of the cell into a more human-readable and concise sentence.
However, users may have varied preferences for the complexity of the generated bullet points.
\tool provides two sliding bar widgets in \tooloutline for adjusting the generation rule.
As shown in \autoref{fig:components} (b3), the left sliding bar is for selecting the number of retrieved cells to be included in a slide; the right one enables users to control the detail level of the generated bullet points.
In addition to the generated bullet points, \tool synchronizes the outline into slide titles, and incorporates outputs (\eg, charts and tables, if available) from selected cells onto the slide.
To provide a meaningful layout, \tool draws inspiration from previous work \cite{wang2019datashot} and adopts the commonly used Parallel layout, which vertically separates each slide into three parts as shown in \autoref{fig:components} (C).
At the top, the slide title is placed. 
Below the title, bullet points are arranged in descending order of relevance scores to the outline item.
Below the bullet points, there are tables and charts, whose sizes are dynamically adjusted to avoid occlusion.
\tool further arranges them from left to right.

%% file: texfiles/5-usagescenario.tex
\section{Usage Scenario}

\begin{figure*}[tb]
  \centering
  \includegraphics[width=0.98\linewidth]{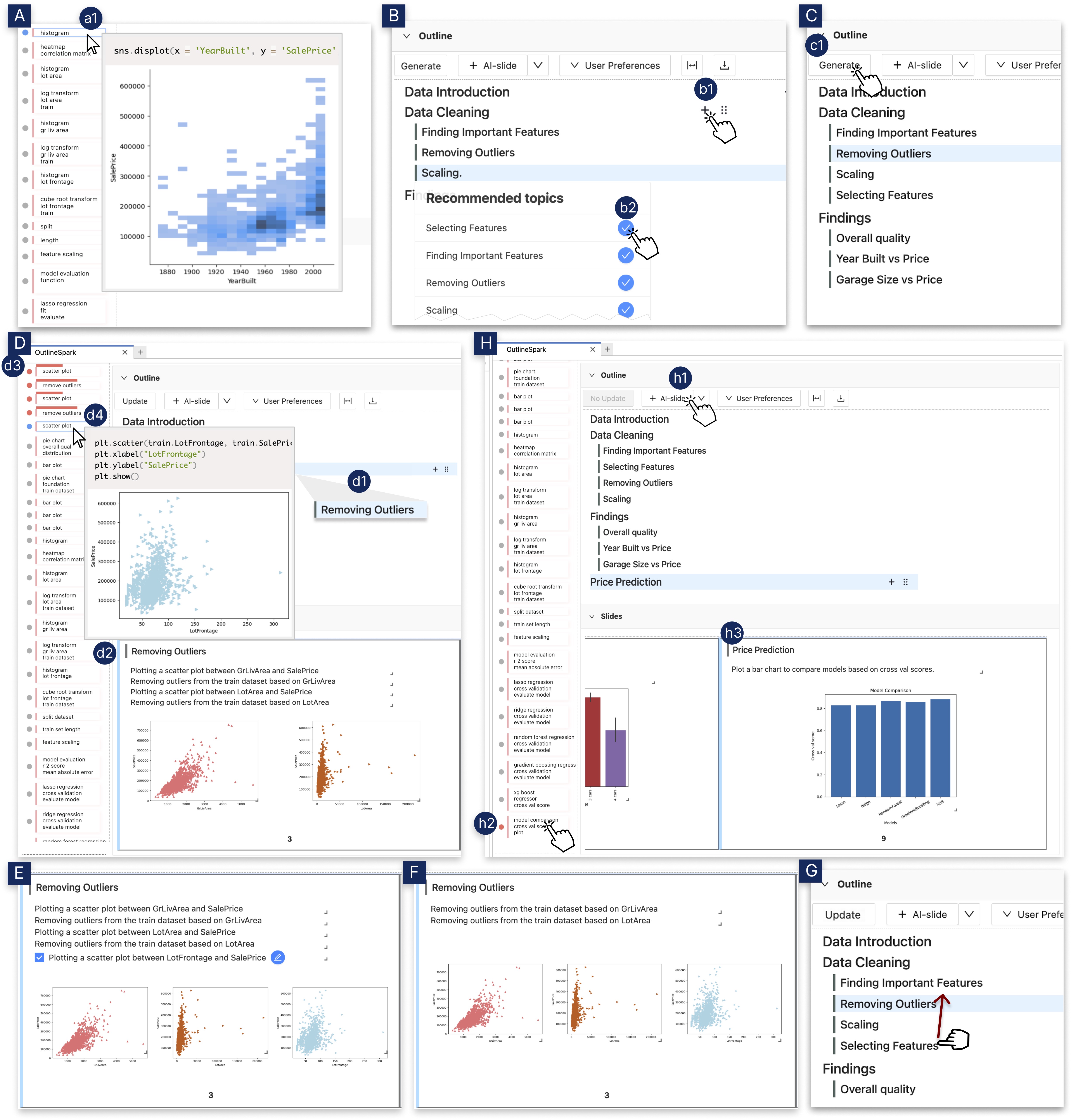} 
  \vspace{-3mm}
  \caption{This figure illustrates a usage scenario in which an analyst, \actor, utilizes \tool to create slides from his computation notebook. In the process, \actor follows these steps: (A) Gain an overview of the notebooks, (B) Draft an outline to guide the slide generation, (C) Instruct \tool to generate the slides, and (D-H) Refine the generated slides.} 
  \Description{The figure presents details of a usage scenario in which an analyst, Andie, utilizes OutlineSpark to create slides from his computation notebook. In the process, Andie follows these steps: (A) Overview the notebooks, (B) Draft an outline to guide the slide generation, (C) Instruct OutlineSpark to generate the slides, and (D-H) Refine the generated slides. Specifically, In (A), Andie activates OutlineSpark and starts with the Notebook Overview. He hovers over a card labeled "a1", representing a specific notebook cell. This action triggers a tooltip that offers details like code snippets and cell outputs.
In (B), Andie constructs an outline in the Outline Panel. He starts by creating three main topics: “Data Introduction”, “Data Cleaning”, and “Findings”. To elaborate on the “Data Cleaning” topic, he clicks a “+” button (marked as “b1”) and adds subtopics manually such as “Finding Important Features”, “Removing Outliers”, and “Scaling”. After that, Andie creates an empty subtopic, presses the spacebar to trigger OutlineSpark's topic recommendation feature, and selects “Selecting Features” (indicated as “b2”), deeming it a suitable addition to his content. He continues this process until he is satisfied with the drafted outline to guide the subsequent slide generation.
In (C), Andie initiates slide creation by clicking the “Generate” button (marked as “c1”). OutlineSpark responds by generating slides for each topic and subtopic at the lowest level of the outline.
In (D), while reviewing the slides, Andie notices the absence of relevant charts in the slide for the topic “Removing Outliers”. He clicks this slide in the Slides Panel, which highlights it with a blue border (marked as "d2") and similarly highlights the corresponding outline item (marked as "d1"). The Notebook Overview auto-scrolls to the relevant cell (marked as "d3"), with this cell and others contributing to the slide being accentuated with pink dots and bars on the left and top, respectively. The width of each pink bar visually represents how relevant a particular cell is to the outlined item. Andie closely examines these cells, including some adjacent ones, and spots a cell displaying a “scatter plot”. Hovering over this cell, a tooltip pops up (identified as “d4”), offering further details about it. After scrutinizing the content of the cell, Andie selects it and clicks the "Update" button, initiating the inclusion of this new cell into the slide.
In (E), the content of the slide is updated with a new bullet point, “Plotting a scatter plot between LotFrontage and SalePrice,” and the scatter plot from the newly selected notebook cell.
In (F), Andie finds some bullet points are redundant and decides to delete them.
In (G), Andie focuses on improving the logical flow of his presentation. He rearranges the outline by dragging and dropping the subtopic “Selecting Features” so that it follows directly after the subtopic “Finding Important Features”.
In (H), Andie notices that he has forgotten to discuss the performance of the models in his outline. As he already knows where the relevant cells are, he directly clicks the “+ AI-slide” button on the Outline Panel (marked as “h1”) and scrolls in Notebook Overview. With a simple click, OutlineSpark automatically scrolls down the notebook window to the actual cell, enabling Andie to view the detailed code. He then selects the cell (marked as “h2”) and clicks the “Generate” button and within a second, a slide with a generated title, bullet points, and charts is rendered on the Slides Panel (marked as “h3”).}
  \label{fig:scenario}
\end{figure*}

In this section, we present a scenario to illustrate how \tool supports structuring and creating slides from computational notebooks.
Suppose \actor is a data analyst working at a real estate company.
His manager tasks him with analyzing a house dataset in order to predict house prices. 
After completing the analysis using Jupyter notebook, \actor wants to create slides to report the analysis results to the manager. 

\actor opens the notebook in JupyterLab and activates \tool from the toolbar.
Then, he puts the two windows side by side to start to create slides based on his notebook.
Next, he has a quick recall of what has been analyzed by browsing the keywords summarized from notebook cells in \toolcode (\autoref{fig:scenario} (A)). 
When \actor hovers over the cards in \toolcode (\autoref{fig:scenario} (a1)), each representing a cell in the notebook, a tooltip appears, providing additional details about the corresponding cell, such as code snippets and the cell's output (e.g., tables and charts). 

Then, \actor proceeds to draft an outline using \tooloutline (\autoref{fig:scenario} (B)). 
The tool distinguishes different levels of outline through variations in font size and indentation. 
He first drafts an outline with three topics: ``Data Introduction'', ``Data Cleaning'', and ``Findings''.
Then he plans to illustrate the topic ``Data Cleaning'' with more sub-topics.
By clicking the ``+'' button (\autoref{fig:scenario} (b1)), he manages to add sub-topics ``Finding Important Features'', ``Removing Outliers'', and ``Scaling'' under the topic.
After that, he would like to check if he missed some critical content.
By focusing on the last sub-topic and pressing the space bar, he gets some recommended topics from \tool and finally selects ``Selecting Features'' which he thinks complements the current content (\autoref{fig:scenario} (b2)).
Following the above practice, he specifies the sub-topics for each topic when it is necessary.
Gradually, an outline forms for generating the slides (\autoref{fig:scenario} (C)).
Next, he clicks the ``Generate'' button (\autoref{fig:scenario} (c1)).
For each topic and sub-topic at the lowest level of the outline, \tool generates a corresponding slide.
\tool selects the top-K (a parameter that can be adjusted by the user) cells relevant to each sub-topic and renders a slide deck in \toolslide (\autoref{fig:scenario} (d2)).
After browsing each slide, he finds most slides present the information he intends to convey.

However, \actor notices that the slide corresponding to the topic ``Removing Outliers'' lacks some related charts (\autoref{fig:scenario} (d2)).
He vaguely remembers noticing three relevant charts while scanning through the notebook in the beginning. 
To check with this, \actor clicks on the slide in \toolslide.
Upon the action, \toolslide highlights the card that represents the slide with a blue left border (\autoref{fig:scenario} (d2)), while \tooloutline highlights the corresponding outline item with a blue background (\autoref{fig:scenario} (d1)). 
Simultaneously, \toolcode automatically scrolls to the first cell used to generate this slide (\autoref{fig:scenario} (d3)).
Additionally, in \toolcode, the cards that represent the cells selected for slide generation are highlighted by pink dots at the left and pink bars at the top.  
The width of a pink bar indicates the relevance of a cell to the outline item.
By examining these cells and several cells nearby them, \actor identifies a cell that displays a ``scatter plot''.
When hovering, a tooltip appears (\autoref{fig:scenario} (d4)), providing additional details about the corresponding cell.
After reviewing the cell, \actor selects it and clicks on ``Update''. 
Within moments, as shown in \autoref{fig:scenario} (E), the content of the slide is updated with a new bullet point, ``Plotting a scatter plot between LotFrontage and SalePrice,'' and the scatter plot from the notebook cell. 
However, \actor doesn't want to include several bullet points, so he deletes them (\autoref{fig:scenario} (F)).

Next, \actor decides to have a final review of the structure of his presentation.
He finds that placing ``Selecting Features'' immediately after ``Finding Important Features'' would be more appropriate.
Thus, he adjusts their orders by the dragging and dropping interaction in \tooloutline (\autoref{fig:scenario} (G)). 
\actor then clicks on the ``Update'' button, prompting \tool to re-order the slides to align with the revised outline. 
Furthermore, he notices that he has forgotten to discuss the performance of the models in his outline.
As he already knows where the relevant cells are, he directly clicks the ``+ AI-slide'' button on \tooloutline (\autoref{fig:scenario} (h1)) and scrolls in \toolcode. 
With a simple click, \tool automatically scrolls down the notebook window to the actual cell, enabling \actor to view the detailed code.
He then selects the cell (\autoref{fig:scenario} (h2)) and clicks the ``Generate'' button. Within a second, a slide with a title, a bullet point, and a chart is rendered on \toolslide (\autoref{fig:scenario} (h3)). 
However, the title doesn't fit his style, he then renamed it to ``Price Prediction'' in \toolslide.
Now he is satisfied with the structure and general content of the slides. Finally, he makes slight modifications to improve the wording and style of the slides.

%% file: texfiles/6-userstudy.tex
\section{User Study} 
To evaluate the usability and effectiveness of \tool, we conducted a user study with 12 participants.
We do not compare our system with a baseline.
That's because the most comparable system is Slide4N~\cite{wang2023slide4n} which generates slides based on cells selected by the users.
However, as we have discussed, our work is not to replace such an interaction but to complement it with the outline-based approach to facilitate slides ideation and streamline the slides ideation and creation process.
Thus, the evaluation of our tool will be primarily reflected by the feedback from participants.
Furthermore, we attempt to understand participants' preferences between the two different interaction approaches, as well as whether different situations would affect their preferences.

\subsection{Participants}
We recruited 12 participants (5 females, 7 males; aged $24.7 \pm 2.6$) through social media and word of mouth (denoted as P1-P12).
They are postgraduate researchers from diverse backgrounds, including human-computer interaction, visualization, recommendation systems, and ocean engineering. 
Participants self-reported their familiarity with Jupyter Notebook with a rating of $5.08 \pm 1.38$, where 1 represents ``No Experience'' and 7 represents ``Expert''.
Moreover, they were familiar with creating slides for presentation, with frequency of 1 to 5 times per week.

\subsection{Task}
In our user study, \rev{inspired by the task in Slide4N \cite{wang2023slide4n} and NB2Slides \cite{nb2slides}}, participants were required to create a slide deck for a 10-minute presentation.
They were told that the target audiences could be business or technical audiences and the presentation content could consist of anything within the notebook that participants deemed important for presentation.
To simulate the realistic scenario wherein participants would create slides following data analysis, they were tasked with familiarizing themselves with the provided notebook first before proceeding to slide creation.
The slide deck was required to contain between 5 and 10 slides, both to gauge participants' proficiency in using \tool and to manage the time constraints of the study.
To better understand users' preferences, during slide creation, participants were required to adopt two types of interactions for guiding slide generation, i.e., creating outlines and selecting cells.

\subsection{Data}
We selected the House Prices Prediction\footnote{https://www.kaggle.com/competitions/house-prices-advanced-regression-techniques} notebook from Kaggle for the experiment, which is commonly employed to assess the effectiveness and usability of tools~\cite{wang2023slide4n, lin2023inksight, nb2slides}.
We removed all markdown cells to avoid the participants being affected by these cells and following the structure of those cells to organize the presentations.
Furthermore, prior research indicates that most data analysts typically avoid documentation due to its time-consuming nature and potential to disrupt analytical flow~\cite{wang2022documentation, lin2023inksight}.
The resulting notebook comprised 42 code cells.

\subsection{Procedure}
\begin{figure*}[tb]
  \centering
  \includegraphics[width=\linewidth]{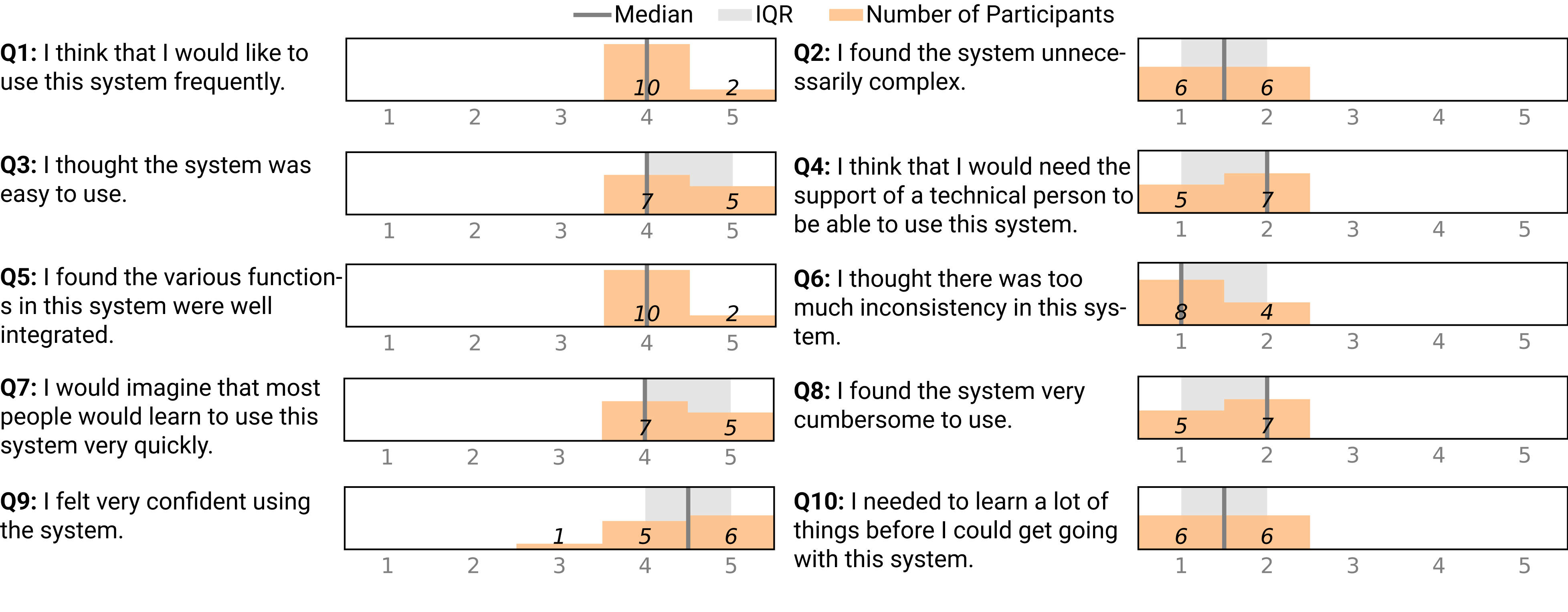} 
  \vspace{-10mm}
  \caption{This figure presents participants' ratings on the usability of \tool on a 5-point Likert Scale (1 = ``strongly disagree'' and 5 = ``strongly agree'').}
  \label{fig:ques-u}
  \Description{The figure presents participants' ratings on the usability of OutlineSpark on a 5-point Likert Scale (1 = "strongly disagree" and 5 = "strongly agree"). The responses to ten different usability questions are presented:
Question 1, "I found the system unnecessarily complex", received 2 ratings of 5 and 10 ratings of 4, leading to a median score of 4 and an Interquartile Range (IQR) of 0.
Question 2, "I think that I would like to use this system frequently", received 6 ratings of 2 and 6 ratings of 1, leading to a median of 1.5 and an IQR of 1.
Question 3, "I thought the system was easy to use," received 5 ratings of 5 and 7 ratings of 4, leading to a median of 4 and an IQR of 1.
Question 4, "I think that I would need the support of a technical person to be able to use this system," received 7 ratings of 2 and 5 ratings of 1, leading to a median of 2 and an IQR of 1.
Question 5, "I found the various functions in this system were well integrated," received 2 ratings of 5 and 10 ratings of 4, leading to a median of 4 and an IQR of 0.
Question 6, "I thought there was too much inconsistency in this system," received 4 ratings of 2 and 8 ratings of 1, leading to a median of 1 and an IQR of 1.
Question 7, "I would imagine that most people would learn to use this system very quickly," received 5 ratings of 5 and 7 ratings of 4, leading to a median of 4 and an IQR of 1.
Question 8, "I found the system very cumbersome to use," received 6 ratings of 2 and 5 ratings of 1, leading to a median of 2 and an IQR of 1.
Question 9, "I felt very confident using the system," received 6 ratings of 5, 5 ratings of 4, and 1 rating of 3, leading to a median of 4.5 and an IQR of 1.
Question 10, "I needed to learn a lot of things before I could get going with this system," received 6 ratings of 2 and 6 ratings of 1, leading to a median of 1.5 and an IQR of 1.}
\end{figure*}

All studies were conducted through one-to-one in-person meetings, each lasting about one hour.
Prior to the user study, we briefly introduced the study procedure and gained consent from participants for video recording the whole process.
The user study was segmented into four distinct phases: a training session, an experiment session, a post-study questionnaire session, and a post-study semi-structured interview.
During the training session, we first briefly introduced the components of \tool and its related interactions.
Participants were then required to interact with \tool to create a slide deck for a sample notebook for around 15 minutes, or until they felt familiar with it.
The sample notebook is about Titanic data with 11 cells.
The subsequent experiment session lasted around 25 minutes, where participants were given the experiment notebook and asked to finish a slide deck with 5-10 slides using \tool.
This session concluded only when participants indicated satisfaction with their created slides, followed by a brief presentation of the slides.
They then proceeded to complete two post-study questionnaires. 
The first was the System Usability Scale (SUS)~\cite{brooke1996sus}, a widely recognized method for assessing tool usability, as shown in \autoref{fig:ques-u}. 
The second was a 7-point Likert scale questionnaire aimed at evaluating the effectiveness of \tool, where 1 signifies ``strongly disagree" and 7 denotes ``strongly agree", as shown in \autoref{fig:ques-e}.
The study ended with a semi-structured interview, focusing on the advantages and disadvantages of the \tool as well as a discussion on two interactions of creating slides, i.e., creating outlines and selecting cells (detailed questions can be found in the supplementary material).
Each participant received a compensation of \$13.50 for completing the user study.

\section{User Study Results}
\begin{figure}[tb]
  \centering
  \includegraphics[width=\linewidth]{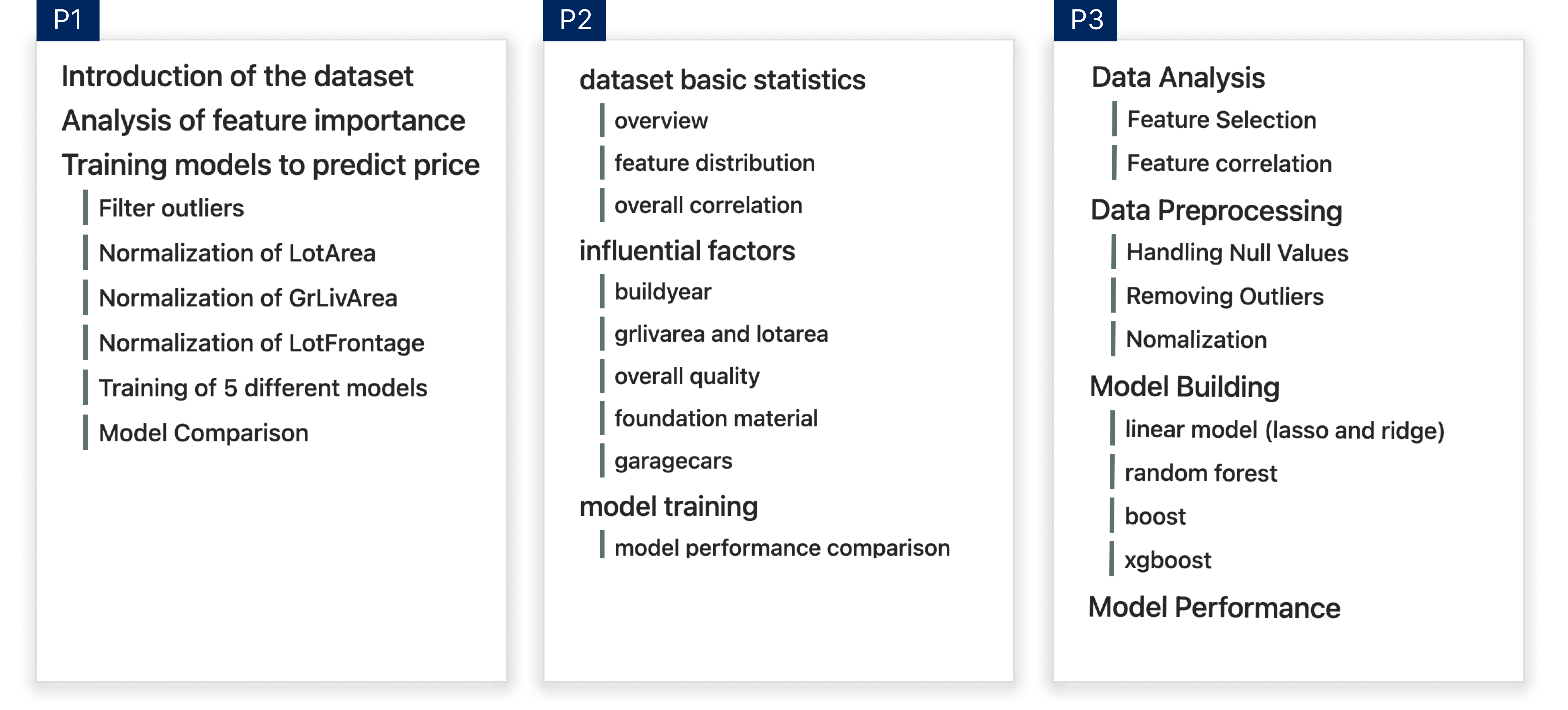} 
  \vspace{-5mm}
  \caption{This figure showcases three sample outlines crafted by participants (P1, P2 and P3) during the user study.
  }
  \label{fig:user-outlines}
  \Description{The figure displays three sample outlines crafted by participants (P1, P2, and P3) in a user study, labeled as figures A, B, and C respectively.
A: Participant P1's Outline - This outline features three main topics: 'Introduction of the dataset', 'Analysis of feature importance', and 'Training models to predict price'. The third topic is further detailed into six subtopics: 'Filter outliers', 'Normalization of LotArea', 'Normalization of GrLivArea', 'Normalization of LotFrontage', 'Training of 5 different models', and 'Model Comparison'.
B: Participant P2's Outline - This outline features three primary topics: 'Dataset basic statistics', 'Influential factors', and 'Model training'. The topic 'Dataset basic statistics' includes subtopics 'Overview', 'Feature distribution', and 'Overall correlation'. The topic 'Influential factors' includes subtopics 'Build year', 'GrLivArea and LotArea', 'Overall quality', 'Foundation material', and 'Garage cars'. The final topic, 'Model training', contains a subtopic on 'Model performance comparison'.
C: Participant P3's Outline - This outline features four main topics: 'Data Analysis', 'Data Preprocessing', 'Model Building', and 'Model Performance'. The topic 'Data Analysis' includes 'Feature Selection' and 'Feature correlation' as subtopics. The topic 'Data Preprocessing' includes 'Handling Null Values', 'Removing Outliers', and 'Normalization' as subtopics. The topic 'Model Building' includes 'Linear model (Lasso and Ridge)', 'Random Forest', 'Boost', and 'XGBoost' as subtopics.}
\end{figure}

In this section,
we first report participants’ questionnaire responses regarding the effectiveness and usability of \tool.
Then, we discuss preferences between outline-based and selection-based slide generation.
The last part is about participants' qualitative feedback on \tool.
As shown in \autoref{fig:user-outlines}, we also present some outlines created by the user study participants.

\subsection{Questionnaire Results} \label{sec:ques}
The quantitative results of our user study reflect participants’ ratings on both the effectiveness and usability of \tool.

\begin{figure*}[tb]
  \centering
  \includegraphics[width=\linewidth]{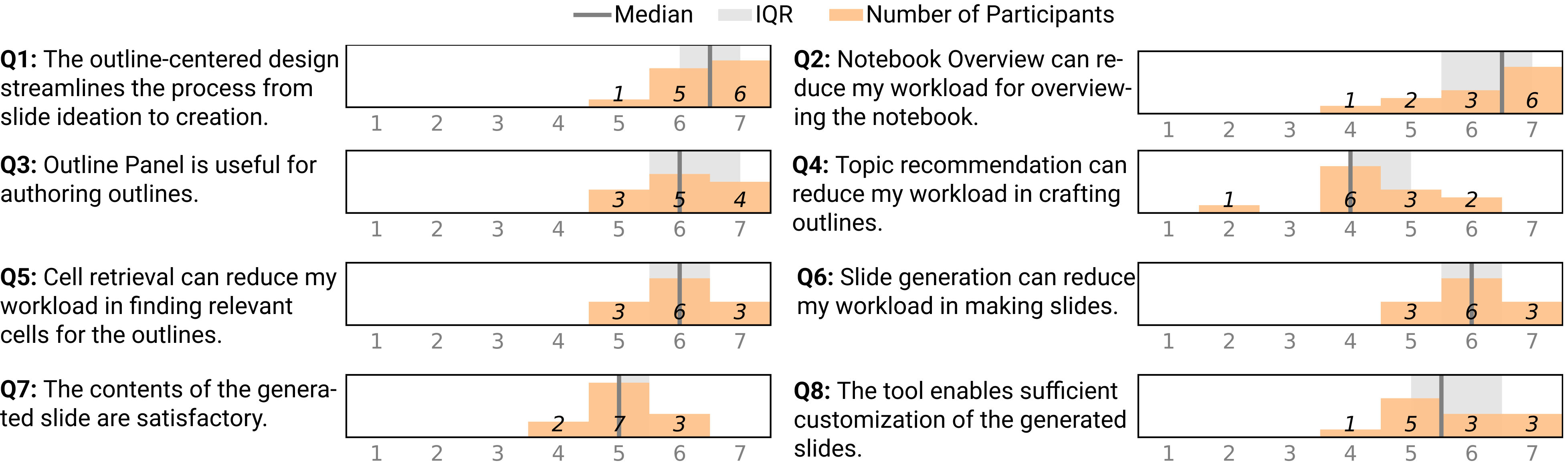} 
  \vspace{-6mm}
  \caption{This figure shows participants' ratings on the effectiveness of \tool on a 7-point Likert Scale (1 = ``strongly disagree'' and 7 = ``strongly agree'').}
  \label{fig:ques-e}
  \Description{The figure presents participants' ratings on the effectiveness of OutlineSpark on a 7-point Likert Scale (1 = “strongly disagree” and 7 = “strongly agree”). The ratings for each question are as follows:
Question 1, "The outline-centered design is useful for streamlining the process from slides ideation to slides creation from computational notebooks," received 1 rating of 5, 5 ratings of 6, and 6 ratings of 7, leading to a median of 6.5 and an IQR of 1.
Question 2, "Notebook Overview can reduce my workload for overviewing the notebook," received 1 rating of 4, 2 ratings of 5, 3 ratings of 6, and 6 ratings of 7, leading to a median of 6.5 and an IQR of 1.5.
Question 3, "Outlines Panel is useful for authoring outlines," received 3 ratings of 5, 5 ratings of 6, and 4 ratings of 7, leading to a median of 6 and an IQR of 1.5.
Question 4, "Topic recommendation can reduce my workload in crafting outlines," received 1 rating of 2, 6 ratings of 4, 3 ratings of 5, and 2 ratings of 6, leading to a median of 4 and an IQR of 1.
Question 5, "Cell retrieval can reduce my workload in finding relevant cells for the outlines," received 3 ratings of 5, 6 ratings of 6, and 3 ratings of 7, leading to a median of 6 and an IQR of 1.
Question 6, "Slide generation can reduce my workload in making slides," received 3 ratings of 5, 6 ratings of 6, and 3 ratings of 7, leading to a median of 6 and an IQR of 1.
Question 7, "The contents of the generated slide are satisfactory," received 2 ratings of 4, 7 ratings of 5, and 3 ratings of 6, leading to a median of 5 and an IQR of 0.5.
Question 8, "The tool enables sufficient customization of the generated slides," received 1 rating of 4, 5 ratings of 5, 3 ratings of 6, and 3 ratings of 7, leading to a median of 5.5 and an IQR of 1.5.}
\end{figure*}

Regarding effectiveness, \autoref{fig:ques-e} depicts the distributions of ratings, as well as the medians (MDs) and interquartile ranges (IQRs) for all participants (detailed ratings can be found in the supplementary material). 
Overall, most of the participants expressed satisfaction with \tool.
Specifically, 11 out of 12 participants highly appreciated the outline-centered design that supported ideation and creation of presentation slides from computational notebooks (Q1), with a median rating of 6.5 (\iqr{1}).
They agreed that the essential interface designs and functions: (1) the outline panel (Q3), (2) the notebook cell retrieval (Q5), (3) the slides generation (Q6, Q7), and (4) the overview of cells (Q2) were useful and met their expectations.
However, the rating for \ctopic (Q4) was relatively lower as most participants gave a neutral score and one participant (P10) gave a negative score of 2.
We will elaborate on the reasons in \autoref{sec:interview} Topic Recommendation.
Furthermore, we noticed that most of the participants weakly agreed that \tool supported sufficient customization of the generated slides (Q8) as \tool can not support rich editing like mature commercial software (e.g., Microsoft PowerPoint) does.

In terms of usability, \tool achieved a score of 85.2, surpassing that of 95\% of applications according to Sauro and Lewis \cite{sauro2016quantifying}.
Notably, during the interview, all participants highly appreciated the seamless integration between the notebook window and the tool, as well as the linking among the three interactive modules. 
As P3 mentioned, \qt{The linking between the slide, the outline, and the notebook cells [in \toolcode] makes the refinement of the generated slides much easier.}

\subsection{Preferences between Outline-based and Selection-based Slide Generation}

\rev{
In this section, we first present our findings by analyzing the slide creation process in the user study.
Then, we present feedback from participants in the interview, regarding the outline-based and selection-based approaches for slide generation. 

\begin{figure*}[tb]
  \centering
  \includegraphics[width=\linewidth]{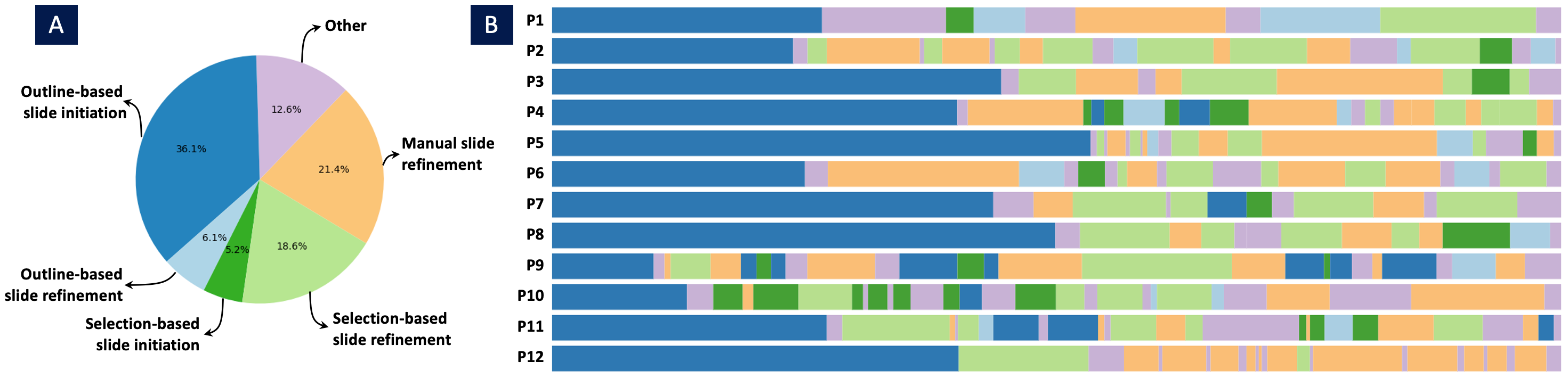} 
  \vspace{-5mm}
  \caption{\rev{The figure showcases: 
  (A) the average proportion of time spent on each activity of slide creation task over 12 participants (P1-P12) in the user study; 
  (B) temporal sequence of activities in 12 separate slide creation sessions, each lasting about 15 to 30 minutes. 
  To facilitate comparison, we normalized the duration of activities within each session for (A) and (B).
  These activities encompassed outline-based slide initiation, outline-based slide refinement, selection-based slide initiation, selection-based slide refinement, manual slide refinement (\ie, editing text and figures on slides), and other (e.g., looking through the user interface).}
  }
  \label{fig:os-comparing-duration}
  \Description{The figure provides a comprehensive analysis of the time allocation and activity sequence in slide creation tasks among 12 participants (P1-P12). It consists of two components: a pie chart detailing the proportion of time spent on specific activities and a stacked bar chart illustrating the sequence and duration of these activities for each participant.
Pie Chart: This chart visualizes the proportion of time each participant spent on different slide creation activities. The activities and their corresponding time percentages and colors are as follows: "Outline-based slide initiation": 36.1\% of the total time, depicted in deep blue. "Outline-based slide refinement": 6.1\%, shown in light blue. "Selection-based slide initiation": 5.2\%, represented in deep green. "Selection-based slide refinement": 18.6\%, indicated by light green. "Manual slide refinement": 21.4\%, portrayed in yellow. "Other": 12.6\%, displayed in purple.
Stacked Bar Chart: This chart illustrates the sequence and proportion of different actions taken during the slide creation activities for each of the 12 participants. For example, Participant P1's sequence of actions is as follows: outline-based slide initiation, other, selection-based slide initiation, outline-based slide refinement, other, manual slide refinement, other, outline-based slide refinement, selection-based slide refinement, and other. Among these activities, the most time-consuming for P1 was outline-based slide initiation.}
\end{figure*}

\begin{figure}[tb]
  \centering
  \includegraphics[width=\linewidth]{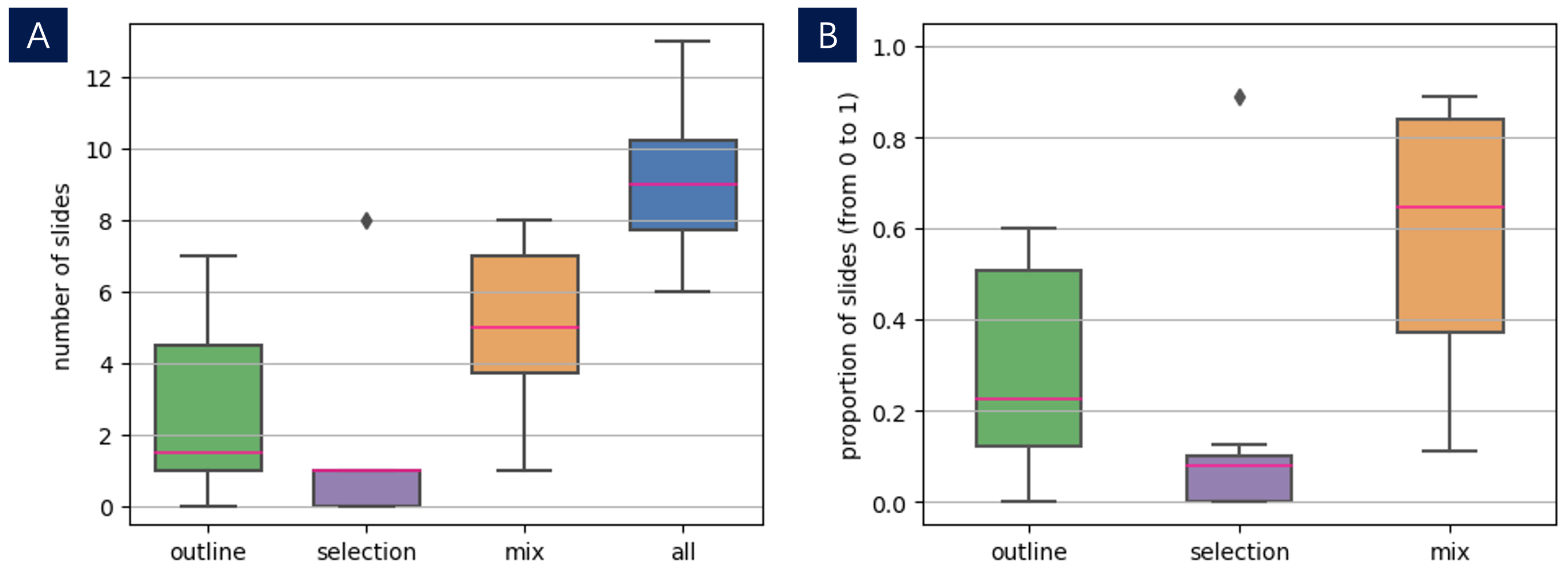}
  \vspace{-5mm}
  \caption{\rev{The figure illustrates the distribution of slide types (\ie, created by outline, selection, or a mix of both) across the 12 decks of slides created by our participants in the user study by boxplots: 
  (A) number of slides;
  (B) proportion of slides.
  The red horizontal lines indicate the median values.}
  }
  \label{fig:os-comparing-slide_no}
  \Description{The figure consists of two charts, View A and View B, both using boxplots to display distribution of slide types (i.e., created by outline, selection, or a mix of both) across 12 decks of slides created by our participants in the user study.
View A displays the number of slides over slide types. Specifically, Slides Created by Outline: The median number is 1.5 slides. The upper quartile is 4.50 slides, and the lower quartile is 1.00 slide. The maximum number is 7 slides, and the minimum is 0 slides. There are no outliers in this category. Slides Created by Selection: The median is 1.0 slide. The upper quartile and lower quartile are both 1.00 slide and 0.00 slide, respectively. The maximum number is 1 slide, and the minimum is 0 slides. There is one outlier at 8 slides. Slides Created by a Mix of Both Methods: The median number is 5.0 slides. The upper quartile is 7.00 slides, and the lower quartile is 3.75 slides. The maximum number is 8 slides, and the minimum is 1 slide. There are no outliers in this category. Total Number of Slides: The median for the total number of slides created is 9.0 slides. The upper quartile is 10.25 slides, and the lower quartile is 7.75 slides. The range is between a minimum of 6 slides and a maximum of 13 slides, with no outliers.
View B displays the proportion of slides analysis over slide types. Specifically, Proportion of Slides Created by Outline: The median proportion is 0.23. The upper quartile is 0.51, and the lower quartile is 0.12. The maximum proportion is 0.60, and the minimum is 0.00, with no outliers. Proportion of Slides Created by Selection: The median proportion is 0.08. The upper quartile is 0.10, and the lower quartile is 0.00. The maximum proportion is 0.13, and the minimum is 0.00, with one outlier at 0.89. Proportion of Slides Created by a Combination of Both Methods: The median proportion is 0.65. The upper quartile is 0.84, and the lower quartile is 0.37. The maximum proportion is 0.89, and the minimum is 0.11, with no outliers.}
\end{figure}

\subsubsection{Characteristics of the Slide Creation Process}

Our analysis of the recorded videos revealed 6 main activities participants engaged with during the slide creation process. 
These activities encompassed outline-based slide initiation, outline-based slide refinement, selection-based slide initiation, selection-based slide refinement, manual slide refinement (\ie, editing text and figures on slides), and other (\eg, looking through the user interface).
We manually labeled the 12 recorded videos based on these activities and their duration, and the results were visualized in \autoref{fig:os-comparing-duration} (more details can be found in the supplementary material).
We initially present an overall summary of our findings before delving into the temporal sequencing.

As shown in \autoref{fig:os-comparing-duration} (A), participants averagely allocated 42.2\% and 23.8\% of their time to outline-based and selection-based slide initiation and refinement, respectively. 
These results underscored the necessity of combining both interaction methods for effective slide creation. 
Specifically, outline-based interaction predominantly facilitated slide initiation and was infrequently employed for refinement. 
Conversely, selection-based interaction was utilized primarily for slide refinement and less commonly for initiation.

Interestingly, we found participants' preferences between outline-based and selection-based approaches were quite consistent throughout the slide creation task.
As shown in \autoref{fig:os-comparing-duration} (B), 10 out of 12 participants initially utilized outlines to plan slides as indicated by the long duration of outline-based slide initiation, followed by a mix of selection, manual editing, and outlines for slide refinement.
Additionally, outline-based slide initiation happened even in later stages, focusing on adding slides through the incorporation of outline items, albeit with shorter duration. 
Selection-based refinement typically preceded manual refinement, signaling participants' inclination to refine slides through cell selection before manual adjustments.
However, no discernible pattern was identified for the timing of outline-based slide refinement.
In the meantime, two exceptions were observed:
P9 did not write a complete outline first to generate all slides, instead, P9 progressively added each item into the outline and generated a slide each time (\autoref{fig:os-comparing-duration} (B)-P9);
P10 predominantly used selection to initiate slides one by one (\autoref{fig:os-comparing-duration} (B)-P10). 
They both mentioned their habits of iteratively organizing materials for slide creation. 
Additionally, P9 stated, \qt{I don't trust LLM is capable of generating high quality slides for my entire outline at one time.}

We further confirm our findings in terms of how slides are created (\ie, by outline, selection, or a mix of both) using boxplots with medians (MDs) marked by red horizontal lines (\autoref{fig:os-comparing-slide_no}).
In general, participants created 6 to 13 pages of slides in the user study (\autoref{fig:os-comparing-slide_no} (A)).
As shown in \autoref{fig:os-comparing-slide_no} (B), the most prevalent approach involved a mix of outline-based and selection-based interactions, constituting a median proportion of 65\% of the slides created in this manner, followed by outline (\md{23\%}) and selection (\md{8\%}).
}

\subsubsection{Feedback on Advantages and Disadvantages}

We present feedback from participants in the interview, regarding the advantages and disadvantages of outline-based and selection-based approaches.

\textbf{Outline-based Slides Generation.}
Participants praised the outline-based slide generation mainly for the following reasons.
First, this approach helps them focus on structuring the slides instead of letting a single slide take too much attention or energy, as highlighted by 9 out of 12 participants (except P2, P6, and P9).
They expressed that the outline-based approach allowed them to focus on the general and essential content of the slides, and free them from diving into a single slide.
As P4 explained: \qt{Outlines serve as the backbone of the story delivered by slides. 
Personally, I prefer to spend more time refining the overall structure to make the entire story more logical. 
If I focus too much on slide details, it prevents me from achieving that.}
Second, it fits their existing slide creation habits (P1, P5, P6, P7 and P12).
As P7 said: \qt{Drafting an outline before creating slides is a common and good practice, which I follow in my day-to-day work.}
Additionally, four participants (P1, P2, P5, P11) noted that the outline-based is more efficient for them.
P1 mentioned that \qt{I just need to organize my story through outlines. \tool would help me find relevant cells from the notebook and convert them into slides.
This process can be quite tedious if done manually. 
Although selecting cells individually to create slides is also helpful, I prefer not to search for these cells throughout the entire notebook.}
P5 also said, \qt{
 When creating slides from notebooks, I have to find the associated cells used for slide creation. 
However, due to the cluttered nature of my notebooks, the cells before or after a specific cell may not necessarily be connected. 
This makes it quite time-consuming for me to locate these cells.
}
Furthermore, participants P4 and P12 expressed that the outline-based approach provided them with a sense of control throughout the slide creation process. 

Besides the advantages, participants pointed out potential improvements of the outline-based slide generation.
First, 4 out of 12 participants (P1, P3, P4, P9) expressed that drafting outlines from scratch could be cumbersome, particularly for casual occasions (P9). 
They suggested the inclusion of a recommended draft outline at the beginning, such as from markdown cells in the notebook, which they can then refine to meet their preferences. 
Second, while most participants found the provided levels (\ie, topic and sub-topic) of outline sufficient, 3 participants suggested including more levels of outline for increased flexibility and granularity.

\textbf{Selection-based Slides Generation.}
When it comes to the advantages of the selection-based approach, there are four reasons.
First, the selection-based approach was recognized as being more accurate compared to the outline-based approach in terms of identifying relevant cells. 
Several participants (P1, P9, P10, P12) said that when selecting cells, the generated slides can better meet their expectations in some cases.
Second, the selection-based approach was considered as a complementary role to the outline-based approach. 
Some participants used this approach mainly to refine the generated slides (P1, P2, P5, and P7), such as dealing with inaccurately retrieved cells, seeking inspiration, or addressing any missing points they noticed in the outline.
Interestingly, three participants (P2, P6, and P10) mentioned that sometimes they knew which cells to use for a slide, but struggled with summarizing them as a topic in the outline. 
In such cases, they found the selection-based approach was beneficial.
Third, the selection-based approach was aligned with some participants' habits of iteratively organizing materials to create slides (P9 and P10).
Rather than setting up outlines from the beginning, they preferred a more iterative approach.
Additionally, this approach was considered more efficient in certain cases. 
P9 and P10 noted that when the notebook or slides were simple, it took less time to select cells compared to typing out the outlines. 

As for disadvantages, the most mentioned one was its general low efficiency.
P12 expressed: \qt{This approach focused on creating slides directly from notebook cells but provided minimal support for slide ideation, which should occupy most of the time when making slides.}
P1 and P3 highlighted the burden of having to manually find and select cells from the notebook to create slides one by one.

\subsection{\tool Streamlines Slide Ideation and Creation from Computational Notebooks} \label{sec:interview}
Next, we report the feedback of our participants regarding some critical functions and designs of \tool.

\textbf{\ckeyword.}
The inclusion of keywords in \toolcode was appreciated by all participants, as it allowed them to gain an overview of the notebook's content without delving into the actual code. 
As P6 said: \qt{I don't want to read the code, I think it's overwhelming when making slides. The keywords really help me grasp a big picture of the notebook quickly.}
Three participants (P2, P4, and P10) echoed with p6.
Furthermore, P12 mentioned that the keywords served as a source of inspiration when drafting outlines. 
Additionally, participants found the keywords helpful in locating specific cells when refining the generated slides, further streamlining the slide creation process.
While participants praised the keywords for effectively summarizing the content of each cell, four participants (P2, P3, P5, and P8) suggested an improvement.
They recommended including X and Y labels in the keywords for charts to help them understand what the chart is about, rather than just displaying keywords such as ``bar plot'', ``histogram'', etc.

\textbf{\ctopic.}
The topic recommendation was not as favored by participants as other functions.
Two participants (P5 and P6) reported that the recommended topics were too specific to the notebook content, while they preferred more general ones.
Furthermore, P2 and P7 suggested that they forgot this function as it is not explicitly indicated in the interface and thus the usage of it is unintuitive.
Participant P10 gave a negative score of 2 and indicated that the recommendations did not consider the context of what P10 had written in the outline. 
However, after reviewing the recorded video, we found it was the latency of LLM that resulted in the failure to present the latest recommendations to P10.
On the other, 5 participants expressed appreciation for the topic recommendation, as it alleviated certain burdens of drafting outlines. 
They turned to the recommendation to get inspiration on what could be added (P7: \qt{I can turn to the recommendation for help when I have no idea what else to present.}) and to check if important content had been missing.
Interestingly, P12 used the recommended topics to refine his outlines. For example, he transformed ``Outliers'' into ``Removing Outliers'' and ``Important Features'' into ``Finding Important Features.'' 
To sum up, we observed that the recommendation feature played an assistant role for participants, which aligned with the design of allowing users to trigger the recommendation only when desired.

\textbf{\cretrieval.}
When asked about retrieving relevant cells from the notebook, 9 out of 12 participants expressed appreciation for its ability to alleviate the manual process of locating cells when creating slides. 
P5 stated, \qt{\tool really helps me find desired cells from the notebook. My notebooks are often long and messy, and I usually have difficulty locating cells.}
Although the retrieved cells were not perfect, participants needed to refer back to check the retrieved cells if the slides didn't meet their expectations, they still valued \tool's assistance in locating cells from the notebook. 
P1 specifically mentioned, \qt{I don't need to look through the whole notebook. Even if the retrieved cells don't match the one I need, it does narrow down the search space.} And this was echoed by P12.

\textbf{\cgeneration.}
In terms of slide generation from notebook cells, all the participants thought it reduced the workload in making slides.
According to P1, \qt{The generated slides match what I want, and I only need to make slight adjustments.} P9 and P11 echoed with P1.
Moreover, P8 and P10 appreciated that \tool saved their time as it automatically arranged titles, bullet points, and charts/tables on slides.
However, participants did provide some suggestions regarding the generated bullet points. 
Both P4 and P10 suggested that while \tool effectively summarized each cell, it would be better to merge similar cells into a single bullet point to avoid repetition. 
Additionally, P2 and P8 expressed the need for automatically extracting insights from the output of cells, particularly for those outputting charts.

\rev{
\section{Slides Quality Assessment} \label{sec:slide_quality_assessment}

We evaluated slide quality based on presenters' self-assessment using Q7 in \autoref{fig:ques-e}; however, these self-impressions may not align with how audiences perceive the slides.
Inspired by Slide4N \cite{wang2023slide4n}, we invited 8 participants to rate the 12 slide decks created by our participants in the user study, followed by a short interview regarding the justifications for ratings.
Ratings covered five aspects: overall satisfaction, clarity of structure, ease of understanding the content, slide layout, and aesthetics.

These participants (4 females, 4 males; aged $25.3 \pm 2.1$) were recruited through social media and word of mouth.
Among them, 6 were postgraduate researchers (referred to as peer reviewers, R1-R6), and 2 were university faculties (referred to as expert reviewers, R7-R8) with extensive experience in presentation slide preparation, delivery, and evaluation.
They came from diverse fields, including data science, human-computer interaction, visualization, machine learning, and computer vision.
The studies were conducted through one-to-one online meetings, each lasting about 30 minutes.
Each participant received a compensation of \$7 for completing the study.

The results of the ratings are illustrated in the boxplots in \autoref{fig:slide-rating} (detailed ratings can be found in the supplementary material).
Overall, both peer and expert reviewers expressed satisfaction with the slides created with \tool, yielding a median rating of 6 for peers and 5 for experts (overall satisfaction).
While peers' ratings on other aspects remained high, experts rated relatively lower.
Upon closer examination, it was observed that R8 gave lower ratings on all aspects, especially on slide layout and aesthetics.
When asked about his concerns, R8 expressed, \qt{
In terms of layout, the text and figures on the slides are detached [when there are multiple figures], requiring additional time for the audiences to match them. 
Aesthetically, the text lacks some highlights, making the slides appear dull over time, and it is challenging for the audience to quickly identify the message the presenter intends to convey. 
Moreover, some bar charts on certain slides are too large, with small accompanying text, resulting in a lack of cohesion in the slides.}
Similarly, two peers (R5 and R6) suggested improving the aesthetic of the generated slides.
As stated by R6: \qt{While aesthetics is not the main consideration when rating the overall satisfaction of slides, certain options can be added to further polish the preset neat slides provided by \tool, such as text highlighting, fonts, and colors.}
To sum up, the generation of slides can be improved in terms of the aesthetics and layout.

}

\begin{figure}[tb]
  \centering
  \includegraphics[width=0.95\linewidth]{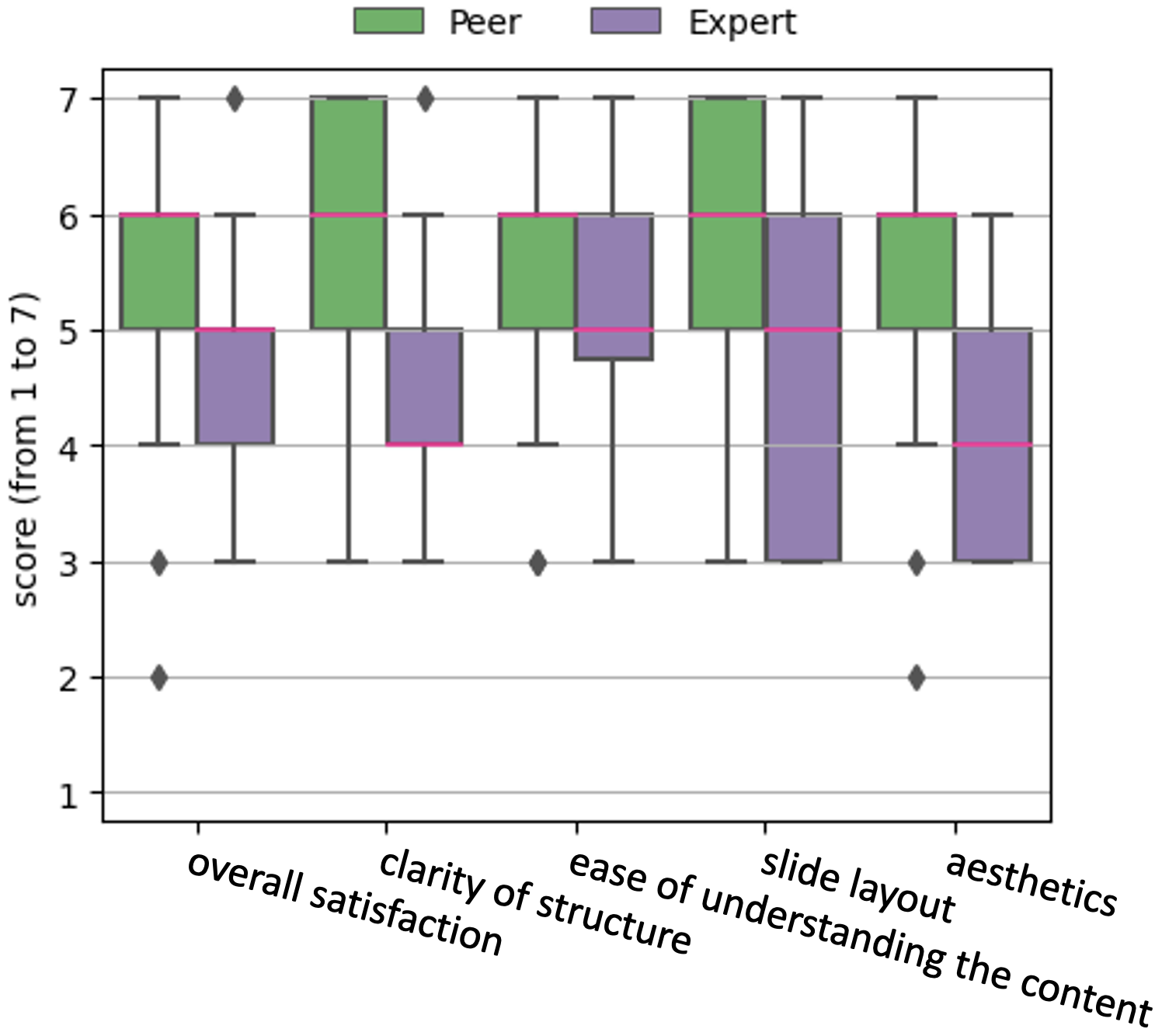}
  \vspace{-1mm}
  \caption{\rev{The figure displays audience ratings for the 12 slides created by \tool in the user study. 
  Ratings are based on a 7-point Likert Scale (1 = "strongly disagree," 7 = "strongly agree") across five aspects.
  The red horizontal lines indicate the median ratings.
  }
  }
  \label{fig:slide-rating}
  \Description{The figure presents audience ratings on five aspects (D1-5) of the 12 slides created by OutlineSpark in the user study. The five aspects include overall satisfaction (D1), clarity of structure (D2), ease of understanding the content (D3), slide layout (D4), and aesthetics (D5), on a 7-point Likert Scale (1 = “strongly disagree” and 7 = “strongly agree”). The red horizontal lines indicate the median ratings.
For overall satisfaction, peers rated a median of 6.0, with upper and lower quartiles at 6.0 and 5.0, respectively, and values ranging from 4.0 to 7.0, including outliers at 2 and 3. Experts rated a median of 5.0, with quartiles at 5.0 and 4.0, and a range from 3.0 to 6.0, with an outlier at 7.
In terms of clarity of structure, the median peer rating was 6.0, with quartiles at 7.0 (upper) and 5.0 (lower), and a range from 3.0 to 7.0, without outliers. Experts gave a median rating of 4.0, with quartiles at 5.0 and 4.0, a range from 3.0 to 6.0, and an outlier at 7.
Regarding the ease of understanding content, peers had a median rating of 6.0, quartiles at 6.0 and 5.0, and values between 4.0 and 7.0, with an outlier at 3. Experts rated a median of 5.0, quartiles at 6.0 and 4.75, and a range from 3.0 to 7.0, without outliers.
For slide layout and aesthetics, peers rated a median of 5.0, with quartiles at 7.0 and 5.0, and values ranging from 3.0 to 7.0, with no outliers. Experts also rated a median of 5.0, with quartiles at 6.0 and 3.0, and a range from 3.0 to 7.0, without outliers.
Lastly, in assessing overall aesthetics, peers rated a median of 4.0, with quartiles at 6.0 and 5.0, and values from 4.0 to 7.0, including outliers at 2 and 3. Experts rated a median of 4.0, quartiles at 5.0 and 3.0, and a range from 3.0 to 6.0, with no outliers.}
\end{figure}

%% file: texfiles/7-v2-discussion.tex
\section{Discussion}
This section first discusses the lessons learned from the research in Section~\ref{sec:design_lessons}.
Then we point out the limitation and potential future work in Section~\ref{sec:limitations}.

\subsection{Design Lessons}\label{sec:design_lessons}
In this section, we present the design lessons learned from our research that can serve as inspiration for the design of future tools.

\textbf{Outline-based slides creation enhances users' ideation and eliminates their workload.}
Crafting outlines before creating slides is a recommended practice \cite{li2023ai, reynolds2011presentation, zanders2018presentation, anholt2010dazzle}.
Building on this concept, we design \tool that supports this workflow for creating slides from computational notebooks.
With \tool, users only need to conceptualize the message they want to convey using outlines, while the tool handles the labor-intensive aspects of slide creation, such as locating relevant cells, distilling key information, and arranging content on slides, to help them to transfer outlines into slides. 
In this way, \tool further bridges the gap between data analysis and presentation.
During our user study, most participants appreciated this workflow, as it aligned with their existing habits and allowed them to plan slides at a high level, resulting in well-structured slides with little effort. 
Consequently, users can dedicate more time to the most critical aspects of the presentation—ideation and planning—rather than getting bogged down with tedious details and tasks, such as locating cells.
We believe future tools can extend the outline-based workflow to other scenarios where individuals gather a set of materials for creating communication materials, such as data articles and videos. 

\textbf{Outline-based and selection-based interaction should be combined for high efficiency.}
\tool provides two types of interactions for users to specify their intent in slide content, i.e., writing an outline or selecting cells.
We found both of them have their own advantages and limitations.
The outline-based approach offers flexibility, facilitating users' ideation and allowing users to freely express their ideas.
However, the comprehensibility of these outlines can sometimes pose challenges for the AI system, resulting in suboptimal retrieval that requires further user refinement.
On the other hand, the selection-based approach offers higher accuracy as users directly choose specific cells for slide generation. 
However, this method can become burdensome when dealing with lengthy and messy notebooks which are common~\cite{head2019managing, rule2018exploration}.
In our user study, we noticed that users often wrote outlines to build their slides rapidly and then leveraged the selection-based approach to fine-tune the generated slides.
They combined both interaction approaches in the task for a more convenient slide creation experience.
Based on the design lesson, we would like to suggest that future tools should consider combining them when designing slide creation tools.

\textbf{\tool facilitates effective collaboration between humans and AI.}
\tool adopts a Human-AI collaboration workflow to facilitate the creation of presentation slides from computational notebooks through outlines. 
The AI assistance in \tool encompasses two aspects: supporting outline creation and transforming outlines into slides.
When using \tool, users first craft outlines with the aid of AI, such as keywords in \toolcode and recommended topics in \tooloutline. 
Subsequently, the AI generates a deck of slides.
Finally users refine them to meet their expectations. 
In our user study, all participants highly appreciated the level of AI assistance in reducing manual work in this tedious task. 
Some even expressed the view that they did not expect full automation in slide creation, even in the time of LLM. 
They believed that the core elements of the task should be controlled by humans, such as the outline in our case, while AI could assume responsibility for handling certain tedious and repetitive tasks.
Considering that individuals have diverse preferences when it comes to slide creation, a promising future direction is to incorporate users' previous slides as input to the AI. 
By leveraging this historical data, AI could learn from users' past slide designs and generate slides that align with their preferences.
This approach would not only generate more personalized slides but also reduce the workload required for slide refinements.

\subsection{Limitations and Future Work}\label{sec:limitations}
In this section, we discuss the limitations and future work of our research, regarding the retrieval accuracy, functionalities, and the evaluations of \tool.

\rev{
\textbf{Retrieval Accuracy.}
With prolonged use of \tool, users may overly trust in AI assistance. 
However, the AI's support may introduce errors, potentially misleading or being overlooked by users.
For instance, the LLM may fail to retrieve cells the users expected as evident in the user study (relevant failed cases can be found in the supplementary material).
It may encounter challenges in comprehending abstract outline items that require additional inference or exhibiting inconsistency in responding to similar or identical queries, leading to retrieved cells that over-represent or under-represent the outlines.
Furthermore, the performance of \tool may be influenced by how organized the notebook is.
An unorganized notebook may have excess or lengthy code cells.
\tool may generate redundant content in the slides when excess cells are retrieved or when only parts of a lengthy cell are what the users desire while \tool retrieves the whole cell.  
To improve the performance of \tool, we plan to investigate methods of cleaning computational notebooks~\cite{head2019managing, kery2019towards, drosos2020wrex}, segmenting a lengthy cell for sub-cell retrieval, and enhancing AI's understanding of users' ways of expressing intentions by learning from historical correct pairs of outlines and desired notebook cells.
}

\textbf{Functionalities.} \label{sec:limitation-func}
The functionalities of \tool can be further extended.
\rev{
First, presentation slides in practice may involve multi-layered structures. 
However, \tool currently provides only two levels of outlines (i.e., topic and sub-topic), which is insufficient in such cases (also pointed out by three participants in the user study). 
Future studies could explore adaptive prompts, allowing \tool to adapt to various levels of cell prioritization based on outline depth and complexity. 
Additionally, conducting an empirical study to explore types of user-crafted outline items would be beneficial.}
Second, \tool currently can't access Python kernel.
It restricts the ability to generate bullet points for chart and table findings that are stored as variables in the kernel, which is considered a point for future improvement by user study participants (P2 and P8).
Following the work on insight generation for charts and tables \cite{li2023notable, lin2023inksight, gong2019table, wang2019datashot}, \tool can be enhanced to generate such bullet points.
Moreover, participants held different opinions regarding the function of recommending potential topics in the outline panel. 
While the recommendation feature was deemed helpful by some participants, others expressed concerns regarding its alignment with their intentions (P5 and P6) and its intuitiveness (P2 and P7).
To cater to different user preferences, future work could consider leveraging insights from users' previous outlines to provide more tailored topic recommendations. 
Additionally, enhancing intuitiveness could be achieved by incorporating a placeholder for blank outline items, such as including a prompt like ``Press space for topic recommendation''. 
\rev{Furthermore, participants R5, R6, and R8 desired the AI-generated slides to be more aesthetically pleasing.
While \tool supports markdown-based editing for styling, it was less convenient than tools like Microsoft PowerPoint \cite{microsoftppt2022}. 
In response, users can export the generated slides as .pptx files for further editing and beautification in PowerPoint. 
Simultaneously, \tool could be enhanced to maintain text-visual layout coherence, thereby reducing users' editing time.}

\textbf{Evaluations.}
In the user study, participants were tasked with creating a slide deck for a notebook using \tool.
The evaluation can be improved regarding four perspectives.
\rev{
First, while we assessed the effectiveness of \tool's main features, we did not evaluate AI assistance concerning trust, accuracy, and perceived cognitive load. 
Examining these aspects could offer additional insights into the tool's overall efficiency.
Second, the quality of the slides could be further evaluated.
While we assessed the quality of slides created with \tool from the perspective of both presenters and audiences, 
a more comprehensive evaluation could be achieved by comparing these AI-generated slides with manually crafted ones (without AI assistance, \eg, PowerPoint), which may provide additional insights into differences in slide quality and efficiency.}
Third, a more long-term evaluation would be beneficial. 
Following previous work \cite{rule2018exploration, wang2022documentation}, we simulated real-world notebooks by removing markdown cells and working with 42 code cells.
However, real-world analysis in computational notebooks can be more complex. 
It would be valuable to gather feedback from users who utilize \tool in their daily work over an extended period. 
Last, we acknowledge that the coverage of participants in our user study was limited. 
We found that the number of participants was relatively small and no data scientists from the industry were involved.
To better assess the value of \tool in real-world settings, future work is needed to conduct a long-term and comparative study with a larger and more diverse pool of participants.

%% file: texfiles/8-conclusion.tex
\section{Conclusion} 
This paper introduced \tool, an interactive and intelligent plugin to streamline the process of creating slides from computational notebooks via outlines.
Utilizing the capabilities of large language models, \tool automates the process by allowing users to define the presentation's structure and content through customized outlines. 
The tool then automatically retrieves relevant notebook cells, extracts essential information, and organizes it within the slides. 
A user study demonstrated that \tool streamlined the slides ideation and creation process from computational notebooks.
Furthermore, participants found that outline-based slide creation aligned well with their existing practices of crafting presentations and allowed them to concentrate on conceptualizing the narrative logic.
In summary, \tool has taken a step forward in bridging the gap between computational notebooks and presentation slides.